\documentclass[12pt]{article}
\pdfoutput=1
\usepackage{amsmath}
\usepackage{jheppub}
\usepackage{graphicx}
\usepackage{slashed}
\usepackage[thinlines]{easytable}
\usepackage{amssymb}
\usepackage{amsthm}
\usepackage{verbatim,graphicx}
\usepackage{mathtools}
\usepackage{color}
\usepackage[all]{xy}
\usepackage{simplewick}
\usepackage{hyperref}
\usepackage{setspace}
\usepackage{array}
\usepackage{tikz}
\usepackage{tkz-euclide}
\usepackage{enumitem}
\usepackage{bbm}
\usepackage{bbold}

\newcounter{yjcc}

\newcounter{nscc}

\newcommand{\Tr}{{\rm Tr}}

\newcommand{\nn}{\nonumber}

\newcommand{\be}{\begin{eqnarray}}
\newcommand{\ee}{\end{eqnarray}}
\newcommand{\vev}[1]{\left\langle #1\right\rangle}

\newcommand{\bmat}{\left ( \begin{array}{cc} }
	\newcommand{\emat}{\end{array} \right ) }

\newcommand{\eqnum}{\refstepcounter{equation}\textup{\tagform@{\theequation}}}

\usetikzlibrary{shapes.misc}

\tikzset{cross/.style={cross out, draw=black, fill=none, minimum size=2*(#1-\pgflinewidth), inner sep=0pt, outer sep=0pt}, cross/.default={2pt}}

\def\Tr{\textrm{Tr}}

\usetikzlibrary{shapes.misc}

\newcommand*{\rdmathspace}[1][h]{%
\if h#1%
\thinmuskip=2mu\medmuskip=1mu\thickmuskip=3mu\fi%
\if m#1%
\thinmuskip=2.3mu\medmuskip=2mu\thickmuskip=3.5mu\fi%
\if s#1%
 \thinmuskip=3mu minus 0.7mu\medmuskip=4mu minus 2mu\thickmuskip=5mu plus 3.5mu minus 1.5mu\fi%
\if d#1%
\thinmuskip=3mu\medmuskip=4mu\thickmuskip=5mu plus 5mu\fi}

\newcommand{\beq}{\begin{equation}}
\newcommand{\beqs}{\begin{equation*}}
\newcommand{\eeq}{\end{equation}}
\newcommand{\eeqs}{\end{equation*}}

\allowdisplaybreaks
\begin{document}

\title{Parisi's hypercube, Fock-space fluxes, and the microscopics of near-AdS$_2$/near-CFT$_1$ duality}
\affiliation{Department of Particle Physics and Astrophysics, Weizmann Institute of Science, Rehovot, Israel}
\author[1]{Micha Berkooz}
 \emailAdd{micha.berkooz@weizmann.ac.il}
\author[2]{Yiyang Jia}
 \emailAdd{yiyang.jia@weizmann.ac.il}
\author[3]{Navot Silberstein}
 \emailAdd{navotsil@gmail.com}
\abstract{
Parisi's hypercube model describes a charged particle hopping on a $d$-dimensional hypercube with disordered background fluxes in the large $d$ limit. It was noted previously [Jia and Verbaarschot, J. High Energy Phys.
11 (2020) 154] that the hypercube model at leading order in $1/d$ has the same spectral density as the double-scaled Sachdev-Ye-Kitaev (DS-SYK) model. In this work we identify the set of observables that have the same  correlation functions as the DS-SYK model, demonstrating that the hypercube model is an equally good microscopic model for near-AdS$_2$/near-CFT$_1$ holography. Unlike the SYK model, the hypercube model is not $p$-local. Rather, we note that the shared feature between the two models is that they both have a large amount of disordered but uniform  fluxes on their Fock-space graphs, and we propose this is a broader characterization of near-CFT$_1$ microscopics.  Moreover, we suggest that the hypercube model can be viewed as the operator growth model of the DS-SYK model. We explain some universality in subleading corrections and relate them to bulk vertices. Finally, we revise a claim made the aforementioned reference about the existence of a spectral gap.
}

\maketitle
\section{Introduction}

Two-dimensional nearly anti-de Sitter (NAdS$_2$) spacetime arises ubiquitously as the near-horizon geometry of near-extremal black holes in higher dimensions. More precisely,  in $D>2$ spacetime dimensions, very often the near-horizon geometry of extremal black holes is of the form AdS$_2\times M^{D-2}$ where $M^{D-2}$ is a compact manifold that describes the shape of the black hole horizon \cite{kunduri2007}. The AdS$_2$ geometry cannot consistently support excitations due to large backreactions. So in order to study excitations we must consider near-extremal black holes and modify the AdS$_2$ geometry to near-AdS$_2$, where we cut off the AdS$_2$ geometry at some finite distance away from the horizon and the cutoff boundary is allowed to fluctuate, consistent with the higher-dimensional flow. The resulting effective theory is Jackiw-Teitelboim (JT) dilaton gravity in two dimensions along with a spatial cutoff \cite{jackiw1985,teitelboim1983,Almheiri:2014cka}, where the dilaton describes the size of the $M^{D-2}$. 

Recently, considerable progress has been made by directly constructing microscopic models for nearly conformal field theory in one dimension (NCFT$_{1}$), most notable of which is the Sachdev-Ye-Kitaev (SYK) model \cite{kitaev2015, maldacena2016, sachdev1993, french1970, bohigas1971}. The SYK model is a system of $N$ Majorana fermions interacting through a $p$-body interaction in which each fermion couples to the rest. At low energy, the model's dynamics matches  that of JT gravity on  NAdS$_2$ spacetime \cite{maldacena2016a}. Apart from its importance in the AdS/CFT correspondence, the SYK model is also important as a solvable model of quantum chaos in $p$-local systems \textit{per se} (and also be realized experimentally \cite{sachdev2023talk}). However, a slightly different large $N$ limit---the double-scaled SYK (DS-SYK) limit  $p, N\to\infty$ with $ \lambda=2p^2/N$ fixed---can be solved exactly in $\lambda$  for all energy scales using the so-called ``chord diagram" technique. The latter technique can also be used to compute correlation functions and also allows for the some reconstruction of the AdS$_2$ dynamics (where it generalizes it to a $q$-deformed AdS$_2$) \cite{Berkooz:2018qkz,Berkooz:2018jqr,Lin_2022, berkooz2022quantum}.

$P$-locality, however, is not the essential ingredient for a quantum mechanical model to have an NAdS$_2$ dual. This is so because there are additional models that are not $p$-local but have the same combinatorial solution. The simplest such example is the hypercube model of Parisi \cite{Parisi:1994jg, Jia_2020}. The model is made out of $d$ qubits, along with a Hamiltonian with interactions that couple together all degrees of freedom in each term (and not just $p\ll N$ of them), albeit in a very specific way. More precisely, we can view
 Parisi's model as a $d$-dimensional hypercubic model where a particle can occupy one of two positions in each lattice direction. The Hamiltonian is just a sum of terms, where each of them is a hopping term in just one direction. However, there are statistically independent nontrivial phases (fluxes) around each plaquette. Compared to the theory without the fluxes, they frustrate the return amplitude of the hopping particle and hence contributes to thermalization of the system.

The aim of this paper is to study further the Parisi model and to clarify which microscopic aspects of the SYK model, or a general quantum mechanical model, are essential and which are not for having an NAdS$_2$/NCFT$_1$ holographic duality, including the gravitational form of quantum chaos. Given these two distinct classes of models (SYK-like and Parisi-like), we will clarify what is in common between the two models, in terms of the dynamics of the Hamiltonian and in terms of a suitable set of observables, and suggest a broader characterization of  NAdS$_2$/NCFT$_1$ microscopics. Enlarging the set of models  and clarifying what is essential for an NAdS$_2$ dual may help build specific examples of AdS$_{D+1}$/CFT$_{D}$ ($D > 1$) that flow to AdS$_2$.

The paper is organized as follows: 
\begin{itemize}
    \item  in section \ref{sec:Model}, we introduce the Parisi model and reinterpret it as a hypercube in the Fock space. 
    
    \item In section \ref{sec:correlations}, we discuss how to solve the model using the moment method; in particular we recapitulate the solution of the spectral density. We then identify the preferred class of observables in this model for which the chord technique applies, which have nice conformal properties at low energies (which demonstrates the emergence of maximal chaos). 
    
    \item In section \ref{sec:importanceOfFluxes}, we discuss why the hypercube model, despite being non-$p$-local,  shares the same phenomenology with the SYK and other $p$-local models. We then propose a unified picture for characterizing NCFT$_1$ microscopics, and speculate how it may arise from higher-dimensional holographic CFTs.    
    \item In section \ref{sec:opGrowth} we discuss how the operator growth of the SYK model may be approximately mapped to a hypercube. 
    
    \item In section \ref{sec:subleading}, we discuss the hypercube moments at subleading order in $1/d$ and demystify a previously observed coincidence with the subleading moments of the sparse SYK model. Furthermore, we discuss how such subleading combinatorics might correspond to an interaction vertex in the  AdS$_2$ bulk. 
    
    \item Finally in section \ref{sec:spectralGap}, we demonstrate that it is unlikely that the model possesses a spectral gap at any nonzero flux, thus revising the claim made in \cite{Jia_2020}.
\end{itemize}
For a short summary of this paper, see companion article \cite{berkooz2023parisis}.

\section{The model and its (re)interpretation}\label{sec:Model}

\subsection{The Parisi hypercube model in symmetric gauge}
We wish to consider the quantum mechanics of a charged particle hopping on the lattice points of a $d$-dimensional hypercube, under the influence of a static uniform background magnetic field. The magnetic field will be drawn from a time-independent probability distribution, namely the magnetic field will be quench disordered.

Let us consider a $d$-dimensional unit hypercube which is centered at the coordinate origin, so that the lattice positions of the hypercube are 
\begin{equation}
    \vec x = (x_1,x_2, \ldots, x_d), \qquad x_i =\pm \frac{1}{2}.
\end{equation}
We will denote  the position eigenstates of the charged particle as $|\vec x\rangle$. We will also use a $\mathbb{C}^2$-qubit representation for the two states in each direction
\begin{equation}\label{eqn:qubitBasis}
 \left|-\frac{1}{2}\right\rangle\rightarrow \begin{pmatrix}
     0\\1
    \end{pmatrix},  \quad       \left|+\frac{1}{2}\right\rangle\rightarrow \begin{pmatrix}
     1\\0
    \end{pmatrix}.
\end{equation}
The Hamiltonian for the hypercube model is defined as
\begin{equation}\label{eqn:symmetricGauge}
\begin{split}
    H&=-\frac{1}{\sqrt{d}}\sum_{\mu=1}^d  D_\mu:=-\frac{1}{\sqrt{d}}\sum_{\mu=1}^d ( T_\mu^++ T_\mu^-)\\
       T^+_\mu &= \prod_{\nu=1, \nu\not=\mu}^d e^{\frac i4 F_{\mu\nu}\sigma^3_\nu} \sigma^+_\mu,\quad \sigma^+_\mu =\frac{\sigma^1_\mu + i \sigma^2_\mu}{2}, \\   T^-_\mu&=(  T^+_\mu)^\dagger,  \qquad F_{\mu\nu}=-F_{\nu\mu},
\end{split}
\end{equation}
where $F_{\mu\nu}$ is a antisymmetric field strength tensor, namely a magnetic field flux on each plaquette (labeled by $[\mu\nu]$). Once two directions are chosen ($[\mu\nu]$) the flux is independent of the rest of the directions in the hypercube. The normalization factor $1/\sqrt{d}$  is chosen for convenience so that the spectral support is finite.

The magnetic fluxes on the plaquettes are drawn from a random disorder distribution $P\left(\{F_{\mu\nu}\}\right)$. For simplicity $F_{\mu\nu}$ (for each $\mu<\nu$) is assumed to be identically and independently distributed (i.i.d.) and to be such that $\vev{\sin F_{\mu\nu}} =0$. Actually, for simplicity, we will assume that it is symmetric in $F_{\mu\nu}$ around 0. Another natural choice is to have $F_{\mu\nu} =0 $ or $\pi$ with any weight. We will see this appearing when we discuss the relation with the SYK model.
The operators $\sigma_\mu^i$ are Pauli matrices $\sigma^i$ acting on the $\mu$th qubit, and the operator $T_\mu^+$ is a parallel transporter that transports the particle to the positive $\mu$ direction while assigning the particle with a phase that lives on the link of the transport.  Hence $H$ is the (negative of) lattice covariant Laplacian (with the removal of a multiple of identity) defined on a hypercube, under the background $F_{\mu\nu}$.   
The parallel transporters satisfy 
 \begin{equation} \label{eqn:VpmAlgebra}
          T_\mu^\pm  T_\nu^\pm =  T_\nu^\pm  T_\mu^\pm e^{i F_{\mu\nu}}, \quad 
     T_\mu^\pm  T_\nu^\mp =  T_\nu^\mp  T_\mu^\pm e^{-i F_{\mu\nu}}
\end{equation}
for $\mu\neq \nu$ and 
\begin{equation}
     (  T_\mu^\pm)^2=0,\quad T_\mu^\pm T_\mu^\mp = \sigma_\mu^\pm \sigma_\mu^\mp.
\end{equation}
These imply
\begin{equation}\label{eq:symmetric-gauge-plaquette}
     T^-_\nu  T^-_\mu  T^+_\nu  T^+_\mu
=e^{-i F_{\mu\nu}}  \sigma^-_\nu \sigma^+_\nu  \sigma^-_\mu  \sigma^+_\mu,
\end{equation} 
 which is our expected holonomy for the particle hopping around an elementary plaquette (a unit square) in the $\mu\nu$ plane under a uniform flux.  in fact, an equivalent but more convenient holonomy formula we shall repeatedly use is 
 \begin{equation}
          D_\nu  D_\mu D_\nu  D_\mu = \cos F_{\mu\nu}-i \sin F_{\mu\nu}\sigma^3_\mu \sigma^3_\nu.
 \end{equation}
 Computations for more complicated loops follow, and we will do such in section \ref{sec:twopt}. The Hamiltonian \eqref{eqn:symmetricGauge} transforms in a rather symmetric way under rotations (see appendix \ref{app:rotations}), and hence we call it the symmetric gauge. This is in contrast with the axial gauge used by Parisi. We now briefly describe the axial gauge for the purpose of bridging the gap with earlier literature, but for the remaining parts of the paper we work exclusively with the symmetric gauge, because correlators are much simpler to compute in this gauge.

\subsection{Relation to Parisi's conventions}
Parisi introduced the hypercube model in the axial gauge where link variables are defined as
\begin{equation}\label{eqn:linkOriginalDef}
U_\mu (\vec x) = e^{i \sum_{\nu=1}^{\mu-1} F_{\mu \nu}x_\nu }, 
\end{equation}
where $\vec{x}$ are coordinates of the hypercube vertices. Note here $\mu$ only denotes positive directions, and the link variables in the negative directions are defined as the inverses of $U_\mu$.  The Hamiltonian can then be defined via matrix elements,
\begin{equation}\label{eqn:hamiAxialGauge}
H_{\vec x, \vec y} = - \frac{1}{\sqrt{d}} \sum_\mu  \left(U_{\mu}(\vec x) \delta_{\vec x +\hat{e}_\mu,\vec y} + U_{\mu}^{-1}(\vec x) \delta_{\vec x -\hat{e}_\mu,\vec y}\right).
\end{equation}
To see it produces the right fluxes, we compute the Wilson loop of an elementary plaquette, the same as what was computed in equation \eqref{eq:symmetric-gauge-plaquette},
\begin{equation}
U^{-1}_\nu(\vec{x})U^{-1}_\mu(\vec{x} +\hat{e}_\nu)U_\nu(\vec{x} +\hat{e}_\mu) U_\mu (\vec x)  =e^{-i F_{\mu \nu} },
\end{equation}
and more complicated loops follow similarly.  Hence we conclude the Hamiltonian \eqref{eqn:hamiAxialGauge} must be related to the Hamiltonian \eqref{eqn:symmetricGauge} by a gauge transformation. The axial gauge Hamiltonian also has a qubit representation which is more complicated \cite{Jia_2020}. Moreover, Parisi studied a particular form of disorder distribution,
\begin{equation}
F_{\mu\nu} = \phi S_{\mu\nu},
\end{equation}
where $\phi$ is a constant in $[0,\pi]$ and $S_{\mu\nu}=\pm 1$ with equal probability, i.e.,
\begin{equation}\label{eqn:originalDisorder}
   P(\{F_{\mu\nu}\}) = \prod_{\mu>\nu} \left[\frac{1}2 \delta(F_{\mu\nu}-\phi)+\frac{1}{2} \delta(F_{\mu\nu}+\phi)\right].
\end{equation}
 In this paper, we will consider both this disorder distribution and more general distributions.   We can compute correlators in the axial gauge, but the results will be  less transparent than those of the symmetric gauge.  We demonstrate some examples in appendix \ref{app:axialGaugeHami}. 

It is worth mentioning on the fly that Parisi's original interest was the second quantized Hamiltonian $\sum_{\vec{x},\vec{y}} \varphi_{\vec{x}}^* H_{\vec{x}, \vec{y}} \varphi_{\vec{y}}$, where $\varphi$ is a complex bosonic field representing the order parameter of the superconducting dots living on the hypercube vertices, so the physics is that of a system of Josephson junctions.  Here we are only concerned with the first quantized Hamiltonian $H_{\vec x, \vec y}$ and a very different kind of physics. The fact that the physics becomes very different should not come as a surprise: this is analogous to removing the spin degrees of freedom from the Sherrington-Kirkpatrick model, upon which only a random matrix remains and all the glassy physics disappears.  The main insight of Parisi is that when $d\to \infty$, a Wilson loop that contributes to $\vev{\Tr H^k}$ simply has the value $(\cos \phi)^A$, where $A$ is the number of elementary plaquettes enclosed by the loop. Parisi also noticed that the $q$-deformed oscillator algebra can be used to solve such a moment problem. These combinatorics are the same as that of the double-scaled SYK model and ensures that the Parisi model has the same NCFT$_1$ physics, which is the inspiration for \cite{Jia_2020} and our current study.

\subsection{The model as a Fock-space hypercube and comments on $p$-locality}
We can (and we will) take an alternative view on what Parisi's hypercube describes.  We  take the Hamiltonian \eqref{eqn:symmetricGauge} as the starting point, and view it as a many-body system of $d$ qubits.  If we represent the basis states as points, and connect two points whenever the corresponding states give a nonzero element for the Hamiltonian, we get back to a hypercube graph. In other words, the hypercube does not live in the real space any more but represents how a state evolves in the Fock space. This is the notion of a Fock-space graph.
In general a Fock-space graph allows us to represent a many-body problem by a single-particle one. For example, the classic reference \cite{altshuler1997} considers the problem of many-body localization-delocalization transition using a complex-SYK-like  Hamiltonian with two- and four-body terms. 
In the occupation number/Fock basis, this Hamiltonian represents a single-particle hopping problem where a single hopping can change the ``Hamming distances'' between basis vectors by 0, 2 or 4 units  and thus defines a graph. The main hope is that  in this manner the many-body localization problem is mapped to a single-particle Anderson localization problem on graphs. However, the resulting problem is still rather complicated.  This often limits the practical usefulness of Fock-space graphs, and in the particular example of  \cite{altshuler1997}, some approximations are made in order to make progress: the Fock-space graph is approximated by a tree graph with a constant and finite node degree (Cayley tree)---see \cite{basko2006} for a more refined discussion and for caveats of this approximation.
On the other hand, the Parisi Hamiltonian in the qubit form \eqref{eqn:symmetricGauge}  describes a system of interacting qubits, but it has a very simple Fock-space graph which is a hypercube. In our case the diverging node degrees of the hypercube play a central role in understanding the chaotic and fast-scrambling behaviors of the model, which is consistent with the heuristics of Sekino and Susskind \cite{sekino2008}.   It is also clear why any kind of tree approximation will obliterate the physics of the Parisi model: without loops one can never see the effect of  fluxes $F_{\mu\nu}$.  Now that the fluxes live in a Hilbert space, it is no longer appropriate to treat them as magnetic fields.  Rather, we find it natural to consider them as Berry curvatures and we will give some justification for this proposal in section \ref{sec:importanceOfFluxes}. 

We stress that our qubit Hamiltonian couples all available qubits together, which is in sharp contrast with $p$-local models such as the SYK model:
\begin{equation}
    H_\text{SYK} = i^{p/2} \sum_{i_1<i_2\cdots <i_p}^N J_{i_1,i_2,\ldots,i_p} \psi_{i_1}\psi_{i_2} \cdots \psi_{i_p},
\end{equation}
or the DS-SYK model in which $p$ is scaled as  $p\sim \sqrt{N}$. The precise definitions for the DS-SYK model will be given in section \ref{sec:importanceOfFluxes}, and here let us just note that the SYK Hamiltonian, though nonlocal, couples $p \ll N$ operators at a time.  
This is called the $p$-locality condition, which is often used for constructing NCFT$_1$ models. The Parisi hypercube model goes beyond this structure.  

We shall remark that a hypercubic Fock-space graph is not a novelty in itself: the transverse Ising model has a hypercubic Fock-space graph as well. The surprise here may be that a Fock-space graph as simple as a hypercube is adequate to produce SYK-type holographic physics. Ultimately we would like to argue that the Fock-space graph geometry (or the $p$-locality condition)  is not crucial, rather it is the large amount of random uniform fluxes that matter. We will come back with a more detailed discussion on this in section \ref{sec:importanceOfFluxes}.

\subsection{Symmetries}\label{subsec:invAndChiralSym}
On the hypercube,  we can implement a parity transformation $\vec x \to -\vec x$ in the qubit representation as 
\begin{equation}\label{eqn:parityOper}
    A = \sigma^1 \otimes \sigma^1 \otimes \cdots \otimes \sigma^1. 
\end{equation}
Since $F_{\mu\nu}$ are invariant under parity transformation, the inversion is a symmetry and indeed one can check that
\begin{equation}
    [A,H]=0
\end{equation}
by using 
\begin{equation}\label{eqn:Vparity}
    A  T_\mu^{\pm} A^
    {-1} =  T_\mu^{\mp}.
\end{equation}
 In the axial gauge, this parity symmetry needs an accompanying gauge transformation to respect the gauge-fixing condition, and was named the ``magnetic inversion symmetry'' in \cite{Jia_2020}.
 
Since the hypercube is bipartite,  it also has a sublattice ``symmetry'' which anticommutes with the Hamiltonian 
\begin{equation}
    \{\Gamma_5, H\}=0,
\end{equation}
where 
\begin{equation}\label{eqn:gamma5def}
    \Gamma_5 = \sigma^3 \otimes \sigma^3\otimes \cdots \otimes \sigma^3.
\end{equation}
This anticommutation relation holds also for $ T_\mu^{\pm}$,
\begin{equation}\label{eqn:Vchiral}
    \{ \Gamma_5,  T_\mu^{\pm} \} =0. 
\end{equation}
Note that equation \eqref{eqn:gamma5def} gives the same operator that represents the sublattice symmetry in the axial gauge, which is not surprising because sublattice symmetry is simply the statement that the hypercube is bipartite, implying that the Hamitonian can be put in a block form where the diagonal blocks are zero.

These symmetries are important for the study of level statistics of the hypercube model, and it was shown in \cite{Jia_2020} that with the disorder \eqref{eqn:originalDisorder} the level statistics follow that of a chiral Gaussian unitary random matrix ensemble, and hence the model is quantum chaotic in the sense of random matrix universality. These symmetries will also play a role in the correlation functions as we shall see in section \ref{sec:correlations}.

\subsection{The velocity operator}\label{subsec:oddParityOp}
It is clear from our definition of position eigenstates that the position operators on the Fock-space hypercube are 
\begin{equation}\label{eqn:positionOperator}
    X_\mu = \frac{\sigma^3_\mu}{2}.
\end{equation}
If we compute the Heisenberg equation of motion for position operators, we obtain the velocity operators
\begin{equation} \label{eqn:DasVelocity}
    \dot X_\mu = i [H, X_\mu] = \frac{1}{\sqrt{d}} V_\mu,
\end{equation}
where
\begin{equation}
V_\mu :=i \left( T_\mu^{+}-  T_\mu^{-}\right).
\end{equation}
This is a parity-odd operator that satisfies 
\begin{align}
      \{ A, V_\mu \} &=0,\\
      \{ \Gamma_5, V_\mu \} &=0.
\end{align}
Among other things, these imply that 
\begin{equation}
    \Tr(H^k V_\mu) = 0,
\end{equation}
or more generally any odd number of insertions of $V_\mu$ in the moments vanish exactly. 
 The introduction of $V_\mu$ is also necessary for closing the algebra of $ D_\mu$ under multiplication and  addition (see appendix \ref{app:EOM}).    There is a gauge transformation that transforms
 \begin{equation}
    ( D_\mu,  V_\mu ) \mapsto (V_\mu, - D_\mu).
 \end{equation}
This is a gauge transformation that assigns a factor of $i$ to all the vertices of one sublattice and leaves the other sublattice invariant. Explicitly, a link variable transforms as
 \begin{equation*}
  e^{i A_\mu (x)} \to   \Omega(x) e^{i A_\mu (x)} \Omega^{-1} (x + \hat e_\mu).
 \end{equation*}
 We can assign $\Omega(x) = i$ and $\Omega(x + \hat e_\mu)=1$.  Note $x$ and $x + \hat e_\mu$ are always on two different sublattices of the hypercube, so this is a consistent assignment that multiplies all forward hoppings by $i$ (and hence backward hoppings by $-i$). This gauge transformation squares to $\Gamma_5$, and it generates a $Z_4$ subgroup of all  gauge transformations. 
 This gauge equivalence motivates us to define the operator
 \begin{equation}
     V := -\frac{1}{\sqrt{d}}\sum_\mu V_\mu
 \end{equation}
which is related to the Hamiltonian $H$ by a gauge transformation, and hence has the same spectrum. Although $V$ is gauge equivalent to $H$, in either given gauge only one of them can be called the Hamiltonian. If we were to call $V$ the Hamiltonian, the parity operator would be $A\Gamma_5$, and $D_\mu$ would be the parity-odd (under this new parity) velocity operator. As we will see in the next section, the velocity operators (with general disorders) serve as one of the good choices for probe operators that give the same correlation functions as those of the double-scaled SYK model.


\section{Moments, correlation functions and maximal chaos}\label{sec:correlations}
The moment problem of computing $\vev{\Tr H^{2k}}$ was solved by \cite{Parisi:1994jg, marinari1995,Cappelli-1998} by considering the combinatorics of loops on the hypercube, and mapping it to a $q$-deformed harmonic oscillator.  The same combinatorial problem can be represented diagrammatically by chord diagrams, and was solved again in the context of double-scaled models first for a quantum spin model \cite{erdos2014}, and later for the DS-SYK model \cite{cotler2016,garcia2017,garcia2018c} following the mathematical work of \cite{ flajolet2000,touchard1952,riordan1975} or using the transfer matrix method of \cite{Berkooz:2018qkz,Berkooz:2018jqr}. In this section we will essentially repeat the same story for  $\vev{\Tr H^{2k}}$, but adapted to the notations we have been using so far. Furthermore, we will study operator insertions in moments $\vev{\Tr \cdots H^{k_3}OH^{k_2}OH^{k_1}}$ and demonstrate it has exactly the same form as the corresponding problem in the DS-SYK model and hence reduce it to a solved problem \cite{Berkooz:2018qkz,Berkooz:2018jqr}.  The key here is the correct identification of an appropriate class of operators, which is one of the main new results of this work. This implies Parisi's hypercube has the same correlation functions as those of the DS-SYK model, including the conformal two-point functions and the out-of-time-ordered four-point function which saturates the chaos bound \cite{maldacena2015}. 
We will favor the chord diagram representation of the combinatorics because it gives a more direct interpretation as   particles propagating in the AdS$_2$ bulk \cite{Berkooz:2018qkz,Berkooz:2018jqr, Lin_2022}. However, the nature of the microscopics of NCFT$_1$ is more transparent in the Fock-space and holonomy language, which we will discuss in due time. 

\subsection{Moments and chord diagrams}
The hypercube model has the moments 
\begin{equation}\label{eqn:momentGeneralHopping}
    2^{-d} \vev{\Tr H^{2k}}  = 2^{-d}\frac{1}{d^k}  \sum_{\mu_1,\ldots \mu_{2k}} \vev{\Tr{D_{\mu_{2k}} \ldots D_{\mu_1}}}.
\end{equation}
Since the trace is a sum over loop amplitudes in the Fock space, each forward hopping must be paired with a backward hopping in the same direction. This means the subscripts $\{\mu_1, \ldots, \mu_{2k}\}$ must form $k$ pairs. Any further coincidence of the $k$ pairs is $1/d$ suppressed, thus we can consider the cases where there are $k$ distinct indices  and there are $(2k-1)!!$ such pairings possible.  When subscripts are distinct, we have the holonomy relation 
\begin{equation}\label{eqn:holonomyParisi}
\begin{split}
      \mathcal{W}_{\mu\nu}&:= D_\nu D_\mu D_\nu D_\mu = \cos F_{\mu\nu}- i \sin F_{\mu\nu} \sigma_\mu^3 \sigma_\nu^3, \\
       q&:=\vev{ \mathcal{W}_{\mu\nu}}= \vev{\cos F_{\mu\nu}}.
\end{split}
\end{equation}
This is equivalent to equation \eqref{eq:symmetric-gauge-plaquette} where the holonomy is computed in terms of $T_\mu^\pm$, but it is more convenient to use $D_\mu$ and we will continue to do so for the rest of this paper. In appendix \ref{app:EOM} we present a more complete list of algebraic relations among $D_\mu$ and $V_\mu$ operators. To compute the moments, we further need  $D_\mu^2 =1$.
The task is then to move the hopping operators around in equation \eqref{eqn:momentGeneralHopping} until the each pair of $D_\mu$
become adjacent to each other and square to identity. We can apply the holonomy relation \eqref{eqn:holonomyParisi} repeatedly, and each time when we have an interlacing ordering of two indices $\mu$ and $\nu$, we get a factor  of $\cos F_{\mu\nu}$ and an imaginary part proportional to $\sin F_{\mu\nu} $. Since we have just argued that all $k$ indices must be distinct at leading order, we are never going to have a $\sin^2 F_{\mu\nu}$ term and all the $\sin F_{\mu\nu}$  ensemble average to zero by assumption; also the $\cos F_{\mu\nu}$ we get must be statistically independent (by definition) and the ensemble average is just a power of $q$, and this power is just the number of interlacing orderings in a sequence of subscripts.  On the hypercube, this power counts the number of elementary plaquettes enclosed by a hopping sequence.  

\begin{figure}
    \centering
    \includegraphics[scale=0.5]{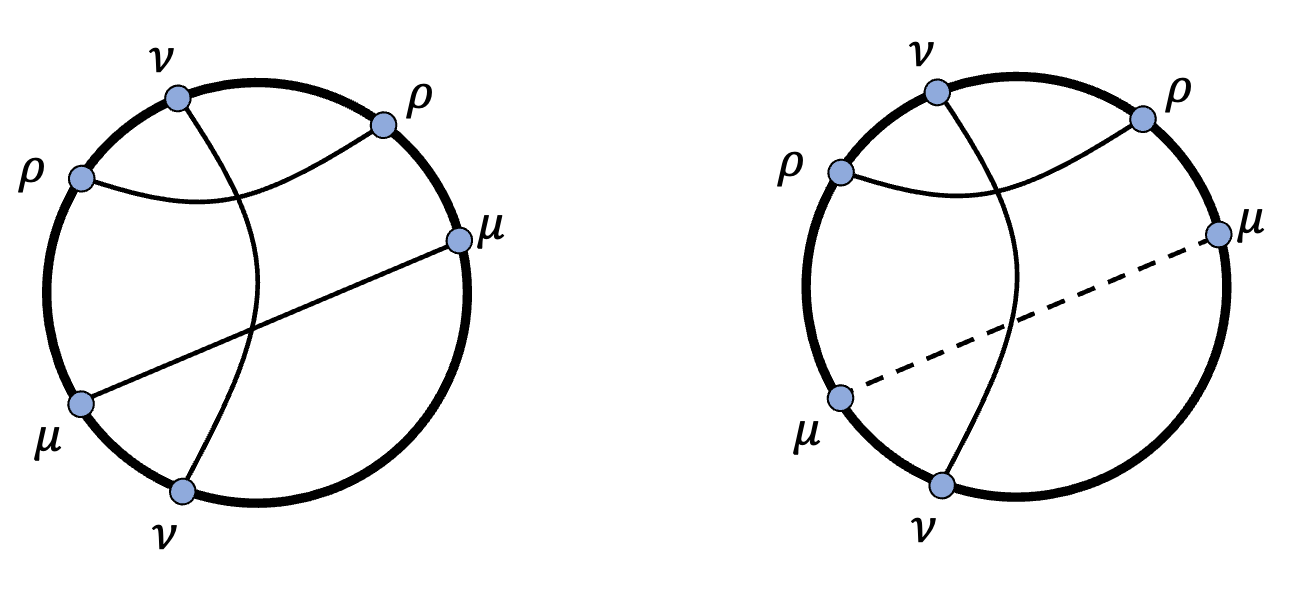}
    \caption{Left: a chord diagram that contributes to the $2^{-d} \vev{\Tr H^6}$, where the repeated indices are summed over. This diagram evaluates to $q^2$ since there are two intersections. Right: a chord diagram that contributes to $2^{-d} \vev{\Tr H^3 O H O}$, where the solid chords represent pairings of hoppings coming from $H$, and the dashed chord represents the pairing of hoppings coming from $O$. This diagram evaluates to $q \tilde q$.}
    \label{fig:parisiChords}
\end{figure}

We can represent the trace diagrammatically as a circle, and each hopping operator as a point on the circle. We connect two points by a chord inside the circle if the corresponding hoppings share a subscript. In this way we obtain what is called chord diagrams, and we illustrate an example in the left panel of figure \ref{fig:parisiChords}. 
Every interlacing ordering of two pairs of subscripts appears as a chord intersection, therefore the general formula for moments is    \begin{equation}\label{eqn:HchordRules}
     2^{-d}\vev{ \Tr H^{2k}}=\sum_{\text{chord diagrams}} q^\text{number of chord intersections}.
\end{equation}
This is identical to the moment formula for the double-scaled SYK model \cite{cotler2016, garcia2017, Berkooz:2018qkz} (and also the double-scaled $p$-spin model \cite{erdos2014}) , and the corresponding spectral density is given by the density function for $q$-Hermite polynomials \cite{ismail1987}:
\begin{align}
  &  \rho(E)= \frac{\Gamma_{q^2}\left(\frac{1}{2}\right)}{\pi \sqrt{1+q}} \left[1- \frac{E^2}{4}(1-q)\right]^\frac{1}{2} \prod_{l=1}^\infty \left[1- \frac{(1-q)q^l E^2}{(1+q^l)^2}\right],\nn
\\
   & \Gamma_{q^2}\left(\frac{1}{2}\right)= \sqrt{1-q^2} \prod_{j=0}^\infty (1-q^{2j+2}) (1-q^{2j+1})^{-1}.
\end{align}
This spectral density behaves as $\sim \sinh \sqrt{E-E_0}$ in the limit 
\begin{equation}\label{eqn:Ncft1Limit}
    q\to 1^-, \quad (-\log q)^{\frac{3}{2}}   \ll  E-E_0 \ll (-\log q)^{\frac{1}{2}}.
\end{equation}
This is sometimes called the triple-scaled limit of the SYK model \cite{cotler2016}, but in this work we will call it the NCFT$_1$ limit to emphasize that it reproduces the NAdS$_2$/NCFT$_1$ phenomenology.

\subsection{Choice of probe operators}\label{sec:randomProbes}

Next we would like to ask what are suitable observables in the theory. For finite $d$ one can discuss any operator on the Hilbert space, but in the limit $d\rightarrow \infty$ not all operators make sense. Mathematically we are interested in operators which survive the large $N$ limit in the sense that their correlation functions are well defined in this limit---this is the starting point for von Neumann algebras. But sometimes, if there is more structure in the theory, we can make a physically motivated choice of a smaller set of observables.

Following \cite{Berkooz:2018qkz}, the AdS/CFT correspondence provides us with a hint on how to select an appropriate set of operators. Suppose that we have some background which starts with a $D$-dimensional AdS space, AdS$_D$, and flows to a near-horizon geometry AdS$_2\times M_{D-2}$, for some compact manifold $M_{D-2}$. Suppose that we model the IR AdS$_2\times M_{D-2}$ in terms of some effective degrees of freedom which are related in some complicated way to the UV degrees of freedom (the details will not matter for us), and suppose that the IR degrees of freedom are described by a Parisi-type model. The operators that we can insert are determined by the UV boundary and hence they will be complicated in terms of the IR degrees of freedom. Inspired by some familiar cases of AdS/CFT, we will refer to them as ``single-trace" operators. We really cannot say much about them beyond their statistical properties. So the problem becomes that of identifying a reasonable statistical class of random observables in the IR theory, i.e., operators acting on the Parisi model's qubits. 

Next note that the Hamiltonian is one such single-trace operator for which the discussion above applies. The rest of the operators that we are interested in are similar single-trace operators and should be of a similar statistical nature when acting on the degrees of freedom with which we model the black hole. For example, in ${\cal N}=4$ super Yang-Mills theory, there is no reason that the local operator $\Tr(X^2)$ and any other local operator $\Tr(X^k)$, or their descendants, would be radically different from each other on the black hole degrees of freedom. 

So our operators will be  similar to the random Hamiltonian, and the simplest construction would be summing over operators analogous to $D_\mu$ or $V_\mu$ but with different fluxes. Namely, we can choose our probes to be 
\begin{equation}\label{eqn:simplestProbesParisi}
    O = -\frac{1}{\sqrt{d}} \sum_\mu \tilde D_\mu  \quad \text{or} \quad  -\frac{1}{\sqrt{d}} \sum_\mu \tilde V_\mu,
\end{equation}
where $\tilde D_\mu$ and $\tilde V_\mu$ are defined in the same manner as $D_\mu$ and $V_\mu$, but with different fluxes $\tilde F_{\mu\nu}$.  To be very explicit, they are constructed out of new parallel transport operators 
\begin{equation}
    \tilde D_\mu =  \tilde T^+_\mu+\tilde T^-_\mu, \quad \tilde V_\mu = i( \tilde T^+_\mu-\tilde T^-_\mu)
\end{equation}
with 
\begin{equation}
    \tilde T^\pm_\mu = \prod_{\nu\neq\mu} e^{\pm\frac{i}{4}\tilde F_{\mu\nu} \sigma^3_\nu} \sigma^\pm_\mu.
\end{equation}
The probe flux  $\tilde F_{\mu\nu}$ is required to satisfy a similar requirement as $F_{\mu\nu}$, namely to be i.i.d. for distinct pairs of $\mu\nu$ and be distributed as an even function in $\tilde F_{\mu\nu}$, however, it may or may not be correlated with $F_{\mu\nu}$. The above two options are by no means exhaustive. More generally, we can consider operators of the form
\begin{equation}
    O_\text{general}=\sum_{\alpha,A} W_{\alpha,A} O_{\alpha,A}
\end{equation}
where $\alpha, A$ together specify a generalized direction in the Fock space,
\begin{equation}
   \alpha=\{\mu_1,..\mu_p\}, A=\{n_1,..,n_p\},\ n_i=\pm,
\end{equation}
$W_{\alpha,A}$ denotes the hopping strengths, and $O_{\alpha,A}$ contains the hopping and phase terms
\begin{equation}\label{eqn:generalProbes}
 O_{\alpha,A}= \left(\prod_{i=1}^p \sigma_{\mu_i}^{n_i} \right) \left(\prod_{\nu=1}^d e^{-i{\tilde f}_{\alpha,\nu }\sigma_\nu^3}\right).
\end{equation}
The random phases $\tilde f_{\alpha, j}$ must be generated from some random uniform fluxes, and  the factor in the second parentheses is an analog of the link variables $U_\mu(\vec x)$ where $\mu$ is replaced by the generalized direction $\alpha$.
So the general random probes are defined by two parameters: how many qubits define a hopping  and the new random phases on the remaining qubits. We recover the simplest probes in \eqref{eqn:simplestProbesParisi} by setting $p=1$ and $\tilde f_{\alpha,\nu }= \tilde F_{\mu\nu}$ and we will focus on this case in the rest of the section.

\subsection{One-point functions}
The one-point function $\langle\Tr e^{-\beta H} O\rangle$ (or any odd-point function) for the choice $O\sim \sum\tilde V_\mu$ simply vanishes 
\begin{equation}
    \Tr (H^{k-1} \tilde V_\mu) =0
\end{equation}
because of the parity and sublattice symmetries discussed in section \ref{subsec:invAndChiralSym} (use sublattice symmetry for odd $k$ and parity symmetry for even $k$). 
The situation with the choice $O \sim \sum \tilde D_\mu$ is more subtle. Consider the  one-point function:
\begin{equation}
    2^{-d} d^{-k/2}\sum_\mu\vev{\Tr (H^{k-1} \tilde D_\mu)}
\end{equation}
with even $k$ (odd $k$ vanishes due to sublattice symmetry). At leading order, a nonzero contribution comes from the pairing of a $ D_\mu$ in one of the $H$'s and the $\tilde  D_\mu$ in $O$. This pairing contribution is given by
\begin{equation}
  2^{-d} d^{-1}\sum_{\mu}\vev{ \Tr \bigl( \sigma_\mu^+\sigma_\mu^-  \prod_{\rho \neq \mu} e^{{i\over 4} (\tilde F_{\mu \rho}-F_{\mu \rho})  \sigma^3_\rho}     + \text{herm.\ conj}\bigr) }=  \vev{ \cos\frac{\tilde F-F}{4}}^{d-1}
\end{equation}
which is to be multiplied by the contribution from the remaining pairings in $\vev{\Tr (H^{k-1} O)}$. Hence the one-point function, and by the same reasoning every odd-point function of $O$, is exponentially suppressed and vanishes at large $d$ as long as 
\begin{equation}
    \vev{ \cos\frac{\tilde F-F}{4}} \neq 1,
\end{equation}
i.e., for any operator other than $H$ (or those differing from $H$ only by a measure zero). Therefore, choosing $\sum \tilde V_\mu$ or $\sum \tilde D_\mu$ makes little difference unless we wish to consider the limit  $\tilde F \to F$, in which case for $\tilde D_\mu$ insertions we need to take the $d\to \infty$ limit first before taking $\tilde F \to F$, whereas for $\tilde V_\mu$ insertions we can set  $\tilde F = F$ from the start.

\subsection{Two-point functions}\label{sec:twopt}
To obtain two-point thermal correlations $\langle\Tr e^{-\beta H} O(\tau) O(0)\rangle$,  we will need to evaluate the two-point moments $\langle\Tr H^{k_2}O H^{k_1}O\rangle$.  We will repeatedly use the relations (see appendix \ref{app:EOM} for more details)
\begin{equation}
    \tilde D_\nu D_\mu \tilde D_\nu D_\mu = \cos \frac{\tilde F_{\mu\nu}+ F_{\mu\nu}}{2} -i\sin\frac{\tilde F_{\mu\nu}+ F_{\mu\nu}}{2} \sigma_\mu^3\sigma_\nu^3,
\end{equation}
or 
\begin{equation}
    \tilde V_\nu D_\mu \tilde V_\nu D_\mu = \cos \frac{\tilde F_{\mu\nu}+ F_{\mu\nu}}{2} +i\sin\frac{\tilde F_{\mu\nu}+ F_{\mu\nu}}{2} \sigma_\mu^3\sigma_\nu^3.
\end{equation}
and that
\begin{equation}\label{eqn:Vsquare}
 D_\mu^2 =\tilde D_\mu^2 = \tilde V_\mu^2 =1.
 \end{equation}
 Also  because we assumed the distribution of $F$ to be an even function, and we have
\begin{equation}
    \vev{\sin \frac{\tilde F + F}{2}} =0.
\end{equation}
Similar to what was discussed in the last section, at large $d$ two $O$ operators must be paired or else there is an exponential suppression.  This means we have a second type of chord which only connects the probes.  We  will call it the $O$ chord (and refer to the Hamiltonian chord the $H$ chord) and draw it using a dashed line, see the right panel of figure \ref{fig:parisiChords} for an example. The chord combinatorics are obtained by essentially the same manipulations used in the last section, and the upshot is that to leading order in $1/d$ we have 
\begin{align} \label{eqn:chordRules}
  &2^{-d}  \vev{
  \Tr H^{k_2}OH^{k_1}O} \nn\\
  = &\sum_{\substack{\text{chord diagrams} \\ \text{with a dased chord}}}  q^{\text{No. of } \text{$H$-$H$ intersections}}\ \tilde q^{\text{No. of } \text{$O$-$H$ intersections}},
\end{align}
where  
\begin{equation}\label{eqn:defsOfqandtildeq}
    q = \vev{\cos F}, \quad \tilde q = \vev{\cos \frac{F+\tilde F}{2}}.
\end{equation}
Again, these are the same chord diagram rules as those of the DS-SYK model \cite{Berkooz:2018qkz, Berkooz:2018jqr}.  Hence we conclude they have the same two-point function, and the conformal dimension of $O$ in the NCFT$_1$ limit [defined by equation \eqref{eqn:Ncft1Limit}] is given by
\begin{equation}\label{eqn:schwarzianOpDim}
    \Delta_O = \frac{\log \tilde q}{\log q}=\alpha \qquad \text{as $\tilde q =q^\alpha \to 1^-$}
\end{equation}
with $\alpha$ fixed. In terms of Fock-space fluxes this gives
\begin{equation}
    \Delta_O = \frac{\vev{(F_{\mu\nu}+\tilde F_{\mu\nu})^2}}{4\vev{F_{\mu\nu}^2}}.
\end{equation}
 As a side remark, we note that the  NCFT$_1$ limit is analogous to taking a continuum limit of the hypercube, reminiscent of that in lattice gauge theories.   This is clearer if we keep a lattice spacing $a$ in our expressions. All the $F_{\mu\nu}$ are replaced by $g a^{2} F_{\mu\nu}$ ($g$ is the gauge coupling), and the continuum limit is where $a\to 0$ with $F_{\mu\nu}$ fixed.  This is the same as taking $F_{\mu\nu} \to 0$ without restoring $a$. Let us keep $a$ explicit for a moment, then the parameters $q$ and $\tilde q$ behave in the continuum limit as 
\begin{equation}
    q \approx 1 -\frac{1}{2} g^2 a^{4} \vev{F^2}, \quad  \tilde q \approx 1 -\frac{1}{2} g^2 a^{4} \vev{\left(\frac{F+\tilde F}{2}\right)^2},
\end{equation}
and the conformal dimension of a probe operator is just a ratio of squared fluxes in dimensionful units.
This continuum limit is not quite the same kind one uses in lattice gauge theories: here we are not keeping any lattice length scale fixed and the number of degrees of freedom does not increase as the limit is taken.  More importantly we interpret the lattice to live in a Hilbert space instead of a real space.

\subsection{Four-point functions}
\begin{figure}
    \centering
    \includegraphics[scale=0.5]{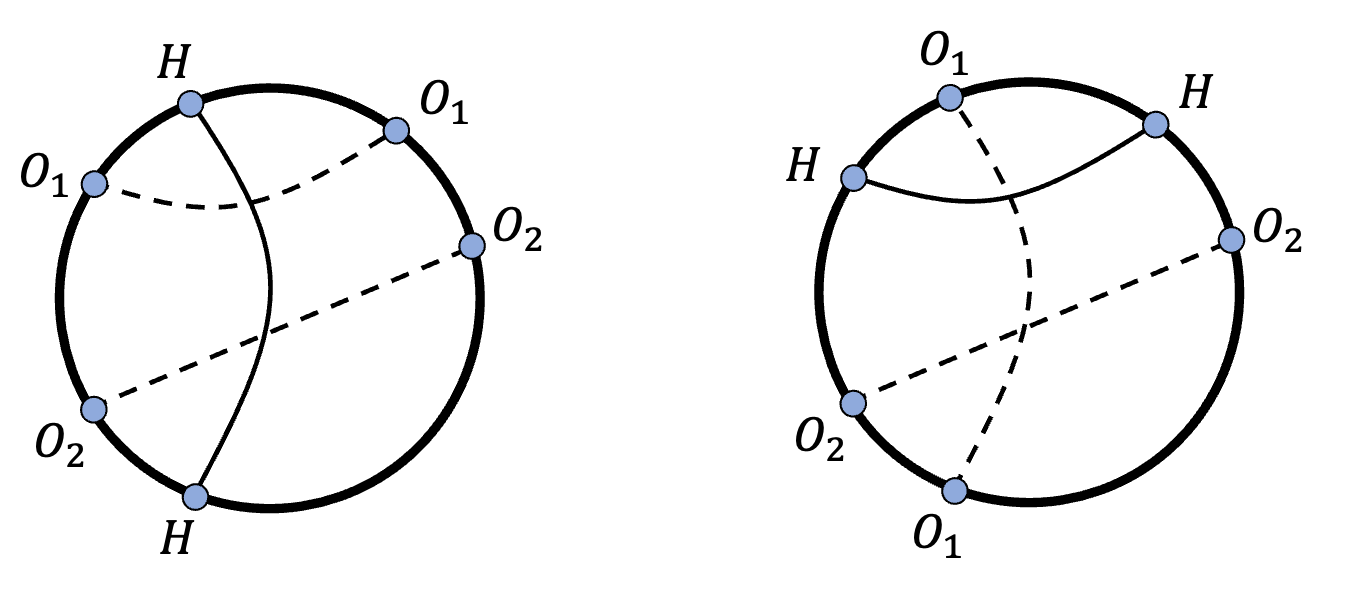}
    \caption{Uncrossed and crossed  four-point insertions. The left figure gives an uncrossed example $2^{-d}\langle \Tr HO_1 O_2HO_2O_1\rangle $ and the right figure gives a crossed example  $2^{-d}\langle \Tr HO_1 H O_2 O_1O_2\rangle$.}
    \label{fig:crossedAndUncrossed}
\end{figure}

We now consider four-point functions. We may define two types of probes $O_1$ and $O_2$ by two probe fluxes $\tilde F^{(1)}$ and $\tilde F^{(2)}$ which may or may not correlate with each other. Following the discussion of the two-point functions we know that at leading order the four-point insertions must be of the form of two pairs, namely $\vev{O_1O_1 O_2O_2}$, $\vev{O_1O_2 O_2O_1}$ and $\vev{O_1O_2 O_1O_2}$.  We will refer to former two as {\it uncrossed} and the last one as {\it crossed} contractions for obvious reasons, and we draw one example for each in figure \ref{fig:crossedAndUncrossed}. If we set $O_1=O_2$ then the result is the sum over all the three contractions, but for simplicity we will keep them distinct.
Note that $O_i$ cannot contract with the Hamiltonian, and intersections between an $O_i$ chord and $H$-chords follow the same rules of equation \eqref{eqn:chordRules}.

\subsubsection{Uncrossed contractions}\label{sec:uncrossed4pt}
For  uncrossed contractions we need to sum over all possible chord diagrams with two dashed chords that are uncrossed, and the summands follow similarly from the discussion of two-point insertions,
\begin{align} \label{eqn:4ptUncrossedChordRules}
  &2^{-d}  \vev{ \Tr H^{k_4}O_2H^{k_1}O_2 H^{k_2}O_1H^{k_1}O_1} \nn\\
  = &\sum_{\substack{\text{chord diagrams with} \\ \text{ two uncrossed dashed chords}}}  q^{\text{No. of } \text{$H$-$H$ inters.}}\ \tilde q_1^{\text{No. of } \text{$O_1$-$H$ inters.}} \tilde q_2^{\text{No. of } \text{$O_2$-$H$ inters.}},
\end{align}
where 
\begin{equation}
    \tilde q_1 = \vev{\cos \frac{F+\tilde F^{(1)}}{2}}, \quad  \tilde q_2 = \vev{\cos \frac{F+\tilde F^{(2)}}{2}}.
\end{equation}
This is  the same rule as that of the DS-SYK model \cite{Berkooz:2018jqr}.
\subsubsection{Crossed contractions and maximal chaos} \label{sec:OTOCscrambling}
One can derive crossed four-point chord diagram rules very similarly, and a new $q$-parameter is needed for the intersection of the $O_1$ and $O_2$ 
\begin{equation}\label{eqn:crossedContribution}
 \tilde q_{12} :=\vev{\cos \frac{\tilde  F^{(1)}+\tilde F^{(2)}}{2}},
\end{equation}
such that
\begin{align} \label{eqn:4ptCrossedChordRules}
  &2^{-d}  \vev{ \Tr H^{k_4}O_2H^{k_1}O_1 H^{k_2}O_2H^{k_1}O_1} \nn\\
  = &\tilde q_{12}\sum_{\substack{\text{chord diagrams with} \\ \text{ two crossed dashed chords}}}  q^{ \text{No. of $H$-$H$ inters.}}\ \tilde q_1^{ \text{No. of $O_1$-$H$ inters.}} \tilde q_2^{ \text{No. of $O_2$-$H$ inters.}},
\end{align}
The main difference with the uncrossed contribution \eqref{eqn:4ptUncrossedChordRules} is  not the appearance of $\tilde q_{12}$, but that the chord diagrams which are summed over have very different structures. 
This is again identical with DS-SYK model chord diagram rules \cite{Berkooz:2018jqr}. Therefore, the Parisi model reproduces the same four-point functions, and consequently its NCFT$_1$ limit also has the same exponentially growing out-of-time-ordered correlator (OTOC) that saturates the chaos bound \cite{Berkooz:2018jqr},
\begin{equation}\label{eqn:OTOCexp}
    \vev{O_1(t_1) O_2(0) O_1(t_2)O_2(0)}_c \sim \lambda^{\text{const}}\exp\left[\frac{2\pi}{\beta} \left(\frac{t_1+t_2}{2}\right)\right],
    \end{equation}
where the subscript $c$ denotes the connected part of the OTOC and $\lambda :=-\log q$. The exponent on $\lambda$ is independent of time or temperature, but can depend on the conformal dimensions of the probes.

We stress that  the factor in front of the exponential growth \eqref{eqn:OTOCexp} is $\lambda$ instead of $1/d$, which implies that the scrambling time is proportional to $\log(\lambda^{-1})$ instead of $\log (\text{entropy})$ \cite{sekino2008}.  The lack of $d$ dependence is expected from a technical perspective: the OTOC is derived from the chord diagrams at leading order of $1/d$, in other words, we are setting $d = \infty$ and it is no longer a parameter. However, we are still owed an intuitive explanation on why $\lambda^{-1}$ serves as a measure of the scrambled degrees of freedoms. In the Parisi model, information is scrambled due to the random phases the hoppings pick up as time evolves.  In the NCFT$_1$ limit $\lambda \to 0$, each hopping only picks up an infinitesimal phase (recall $\lambda \sim \vev{F^2}$) and it is reasonable to expect that the scrambling only happens when the cumulative random phase reaches $O(1)$. Hence as $\lambda$ gets smaller, more hopping steps are needed to achieve this; let us also recall that the large $d$ limit requires each hopping to be along a distinct direction on the hypercube, so more directions---equivalently, more qubits---must be explored in order to get an $O(1)$ cumulative phase. Therefore,  $\lambda^{-1}$ measures the number of degrees of freedom an initial state must explore before the information gets scrambled.


\section{The importance of random uniform fluxes}\label{sec:importanceOfFluxes}
We have demonstrated that Parisi's hypercube  model has the correct NAdS$_2$/NCFT$_1$ physics by showing that it has the same correlation functions as those of the DS-SYK model. We achieved this by showing that at leading order it has the same chord combinatorics as those of the DS-SYK model. However, the hypercube Hamiltonian looks nothing like the SYK Hamiltonian, in particular it is not $p$-local. This is in contrast to the fact that all models that exhibit NAdS$_2$/NCFT$_1$ physics discovered until now were $p$-local. The question is then what exactly do the two types of models have in common.  In this section we try to address this question, and the answer will be a simple characterization of  NCFT$_1$ microscopics. We will further speculate how such microscopics can arise from higher-dimensional holography.
\subsection{The SYK model is also a model of Fock-space fluxes}\label{sec:sykchordReview}
The SYK Hamiltonian of $N$ Majorana fermions can be written as 
\begin{align}\label{eqn:SYKhamiDef}
       H= \Sigma_I J_I \Psi_I
\end{align}
where $I$ is a multi-index of length $p$ ($p$ is an even integer),
\begin{equation}
    I = (i_1, i_2, \ldots, i_p), \quad  1\leq i_1<i_2<\cdots<i_p \leq N,
\end{equation}
so the Hamiltonian is a sum over $\binom{N}{p}$ terms.  Moreover $J_I$ are Gaussian random variables that are independently and identically distributed, with the variance 
\begin{equation}
    \vev{J_I^2} = \binom{N}{p}^{-1}.
\end{equation}
The operators $\Psi_I$ are defined as 
\begin{equation}
    \Psi_I = i^{p/2} \psi_{i_1}\psi_{i_2} \cdots \psi_{i_p},
\end{equation}
where $\psi_{i}$ are Majorana fermions satisfying 
\begin{equation}
    \{\psi_{i},\psi_{j}\} =2\delta_{ij}.
\end{equation}
The operators $\Psi_I$ can be viewed as a linear combination of hoppings much like the $D_\mu$ operators in the hypercube model, and the subscript $I$ effectively specifies a direction in the Fock space. Specifically, we write each Majorana fermion $\psi$ as a linear combination of complex fermions. The latter are fermionic ladder operators that can flip qubits and thus define hoppings in the Fock space \cite{Altland:2017eao, altshuler1997}. In the Majorana  SYK model there would be $N/2$ qubits,  and  the hopping is restricted by Hamming distances  $0, 2, 4, \ldots, p$.  

Similar to hoppings on the hypercube, an elementary holonomy in the SYK Fock space is given by
\begin{equation}\label{eq:SYKPhas}
\mathcal{W}_{IK}:=\Psi_K \Psi_I  \Psi_K \Psi_I =(-1)^{|I\cap K|},
\end{equation}
where $|I\cap K|$ is the cardinality of the set intersection of $I$ and $K$.  Such holonomy gives rise to uniform fluxes in the Fock space, where by this we mean that the right-hand side of equation \eqref{eq:SYKPhas} is constant under changing the location of the loop (in this case it is proportional to the identity operator). So overall, the phase of a loop in the Fock space only depends on the sequence of hopping directions that it defines, but does not depend on where the loop is located.  Comparing with the holonomy formula \eqref{eqn:holonomyParisi}, we see that the fluxes of the SYK model are $0$ or $\pi$. However, since $|I\cap K|$ is Poisson distributed with mean $p^2/N$, then the average of $(-1)^{I\cap K}$ [over pairs of index sets $(I,K)$], i.e., the average holonomy, is between 0 and 1 [equation \eqref{eqn:SYKaveholonomy}] and plays the same role as $q$ above and will be important below.

After averaging over the Gaussian couplings, the moments for the SYK model become
\begin{equation}\label{eqn:wick}
    2^{-N/2}\vev{\Tr H^{2k}} =  2^{-N/2} \binom{N}{p}^{-k} \sum_{\text{paired $I$'s}}\Tr(\Psi_{I_1}\Psi_{I_{2}}\ldots),
\end{equation} 
where there are $(2k-1)!!$ pairings in total, and for each pairing  an Einstein summation convention is assumed on the paired subscripts. Hence we can  represent the SYK moments by chord diagrams as well, in the same way as in figure \ref{fig:parisiChords}.  By repeatedly using the SYK holonomy \eqref{eq:SYKPhas} and that $\Psi_I^2 =1$, for a given chord diagram $G$ we have
\begin{equation}
     2^{-N/2}\vev{\Tr H^{2k}} = \binom{N}{p}^{-k}\sum_{I_1,\ldots,I_k} (-1)^{c(G)},
\end{equation} 
where $c(G)$ is the sum of  all $|I_i\cap I_j|$ as long as the $I_i$ chord and $I_j$ chord intersect in the chord diagram $G$.
Thus, the SYK moment is also a sum over Fock space holonomies whose phases are generated by uniform fluxes (of $0$ and $\pi$), just as in the hypercube model. These fluxes are effectively random since $|I_i\cap I_j|$ can be viewed as a number sampled from a long list of numbers.   However, the analogy with the hypercube model is not yet complete: in the fixed $p$ SYK model each operator is so short that there is generically no nontrivial holonomy ($|I_i\cap I_j|=0$ with probability 1), and one then needs to go to subleading orders in $1/N$ to retrieve nontrivial physics \cite{garcia2018c,Jia_2018}.  A complete analogy is achieved by going to the double-scaled limit,
\begin{equation}\label{eqn:doubleScaling}
    \lambda=\frac{2p^2}{N}\ \text{fixed}, \quad p,N \to \infty. 
\end{equation}
In this limit, $|I_i\cap I_j|$  has a finite probability of being nonempty.  Moreover, set intersections $I_i\cap I_j$ are independently random variables where triple (and higher) set intersections are negligible, and hence the chord intersections factorize. The average of an elementary holonomy in this limit becomes
\begin{equation}\label{eqn:SYKaveholonomy}
  \binom{N}{p}^{-2} \sum_{I,K} (-1)^{|I\cap K|} \to q = e^{-\lambda}.
\end{equation}
Note here the averaging is an effective one, that is, each plaquette has a unique value for its holonomy, but we average over holonomies on all plaquettes. In the hypercube model, the average is over an \textit{a priori} distribution of fluxes.   Hence we arrive at
\begin{equation}
       2^{-N/2}\vev{\Tr H^{2l}}_\text{DS-SYK} = \sum_\text{chord diagrams} q^{\text{No. of chord intersections}},
\end{equation}
which is the same as the hypercube result.  By the same random probe argument as  in section \ref{sec:randomProbes}, the DS-SYK random probes have the form
\begin{align}\label{eqn:SYKprobes}
       O= \Sigma_{\tilde I} \tilde J_{\tilde I} \Psi_{\tilde I}
\end{align}
where $\tilde I$ is a multi-index of length $\tilde p$ ($\tilde p$ is an even integer that scales as $\tilde p\sim \sqrt{N}$), and $\tilde J_{\tilde I} $ is a Gaussian disorder independent of $J_I$. The route toward the chord diagram description of the correlation functions is the same as the hypercube model, that is, by considering mixed holonomies 
\begin{equation}
        \Psi_{K}\Psi_{\tilde I}\Psi_{K}\Psi_{\tilde I} = (-1)^{|\tilde I \cap K|}.
\end{equation}
The upshot \cite{Berkooz:2018qkz, Berkooz:2018jqr} is the same chord combinatorics for correlation functions as those of the hypercube model, including equations \eqref{eqn:chordRules}, \eqref{eqn:4ptUncrossedChordRules} and \eqref{eqn:4ptCrossedChordRules} for two- and four-point functions. For the DS-SYK model the extra $q$-parameters are
\begin{equation}
    \tilde q_1 = e^{-\frac{2 \tilde p_1 p}{N^2}},\quad   \tilde q_2 = e^{-\frac{2 \tilde p_2 p}{N^2}},\quad   \tilde q_{12} = e^{-\frac{2 \tilde p_1 \tilde p_2}{N^2}}.
\end{equation}
We present a comparison of the $q$-parameters in table \ref{tab:compareq}. From this comparison, it is clear that the  number of fermions  ($p$ or $\tilde p$) in the operators $\Psi$  measures the strength of the flux it carries in the Fock space, at least in the NCFT$_1$ limit.  
\begin{table}
    \centering
    \begin{TAB}(b,1cm,1cm,1cm,1cm){|c|c|c|}{|c|c|c|c|c|}
            &  Hypercube & DS-SYK\\ 
        $q$ & $\langle\cos F\rangle$  & $\exp(-2p^2/N)$ \\ 
        $\tilde q$ &  $\langle\cos [(F+\tilde F)/2]\rangle$ &  $\exp(-2p \tilde p/N)$  \\ 
        $\tilde q_{12}$& $\langle\cos [(\tilde F^{(1)}+\tilde F^{(2)})/2]\rangle$ & $\exp(-2\tilde p_1 \tilde p_2/N)$ \\ 
        $\Delta_O$&  $\langle(F+\tilde F)^2\rangle/(4\langle F^2\rangle)$ & $\tilde p /p$ \\
    \end{TAB}
    \caption{Comparison of the $q$-parameters and the conformal dimensions in the hypercube and DS-SYK models.}
    \label{tab:compareq}
\end{table}

\subsection{$p$-local operators as Fock-space flux generators}
The discussion regarding the DS-SYK model in the last section applies to general $p$-local systems. By a $p$-local system, we mean
\begin{itemize}
    \item the Hilbert space is made out of $N$ copies of some basic Hilbert space  $\bigotimes_{i=1}^N \mathcal{H}_i$. For fermionic models it is more appropriate to think of  $\bigoplus_{k=0}^N\left(\bigwedge_{i=1}^k \mathcal{H}_i\right)$.
    \item On each ${\cal H}_i$ there is a family of noncommuting operators (acting irreducibly on ${\cal H}_i$, or else we split each ${\cal H}_i$ to its components). Denote these operators by $s_i^a,\ i=1\ldots N,\ a=1\ldots L.$ 
    \item Denote by $I$ a subset of length $p$ of ordered distinct integers between 1 and $N$, and by $A$ a vector of length  $p$ of integers each taking values between 1 and $L$. Define monomials of the form
    \begin{equation}
    \Sigma_{I,A} = \Pi_{i\in I} s_{i}^{A_i}.
    \end{equation}
    The Hamiltonian is then given by a sum over $\Sigma_{I,A}$ for all $(I,A)$ with random  coefficients with zero mean and finite second moment.
    \item The double-scaled limit is $p^2/N$ fixed with $p,N\rightarrow\infty$ (for fixed $L$, otherwise $L$ can be absorbed into the double-scaled limit as well).
\end{itemize}
The moments of all such Hamiltonians are essentially the same in the double-scaled limit and are captured by chord diagrams. A well-studied example in this class other than the DS-SYK model is the double-scaled $p$-spin model of Erd{\H{o}}s and Schr\"oder \cite{erdos2014,Berkooz:2018qkz},
\begin{equation}\label{eqn:ESmodel}
    H_{ES} = \sum_{i_1,
\ldots i_p}\sum_{a_1,
\ldots a_p} J_{i_1,\ldots,i_p}^{a_1,\ldots a_p} \sigma_{i_1}^{a_1} \cdots \sigma_{i_p}^{a_p},
\end{equation}
where $ \sigma_{i}^{a}$ is one of the three ($a=1,2,3$) Pauli matrices on the $i$th qubit.  

For a double-scaled model, a pair of monomials $\Sigma_{I,A}$ and $\Sigma_{I',A'}$ do not generically commute, hence the model is highly frustrated. However, the operator that encodes the holonomy
\begin{equation}
    \mathcal{W}_{(I,A), (I',A')}:= \Sigma_{I',A'}\Sigma_{I,A}\Sigma_{I',A'}\Sigma_{I,A}
\end{equation}
 is an operator acting on a finite number of ${\cal H}_i$ in the intersection $I\cap I'$, with probability 1 in the thermodynamic limit. 
In particular, $|I\cap I'|$ is Poisson distributed with finite mean determined by $p^2/N$. Having a finite mean is important because this would allow the $q$-parameter to be tunable, or else $q$ will always be 1. 

It is also important that the triple (and higher) intersections $I\cap I'\cap I''$ are negligible, i.e., it is empty with probability 1. This ensures that almost always
\begin{equation}\label{eqn:uniformFluxOperator}
     [\mathcal{W}_{(I,A), (I',A')}, \Sigma_{I'',A''}] = 0.
\end{equation}
Since $\Sigma_{I,A}$ are the basic hopping operators in the Fock space, equation \eqref{eqn:uniformFluxOperator} expresses the fact that the holonomies are generated by uniform fluxes in the Fock space.   From this perspective, $p$-local models  are convenient ways---but not the only ways---of generating a large amount of random uniform fluxes in the Fock space, and the double scaling ensures the holonomies to have a tunable average while being statistically independent.

The models we have discussed so far all have the accidental property that (forward plus backward) hopping operators square to identity:  $D_\mu^2=1$ for the hypercube and $\Sigma_{I,A}^2 =1$ for the SYK model and the $p$-spin model \eqref{eqn:ESmodel}.  This is not essential to the physics. In fact, if we replace the Pauli matrices in the hypercube model or the $p$-spin model by a higher-spin representation of $su(2)$, say spin-3/2 representation, then  the hoppings no longer square to identity, but the chord diagram combinatorics remain the same.  In the Fock space of the hypercube model, this amounts to having a hypercubic lattice with four lattice sites in each direction, and the hopping strengths can vary, but the properties of the fluxes remain unchanged.

\subsection{A  characterization of  NAdS$_2$/NCFT$_1$ microscopics}\label{sec:microscopics}
To summarize, we get $q$-combinatorics for chord diagrams and therefore NAdS$_2$/NCFT$_1$ physics, if a model has Fock-space fluxes that are both
\begin{enumerate}
    \item uniform and quench-disordered, and
    \item independently and identically distributed (i.i.d.) on different (nonparallel) plaquettes of the Fock-space graph, with  a real and tunable average holonomy.
\end{enumerate}
The NAdS$_2$/NCFT$_1$ physics arises when the variance of the fluxes is tuned to zero after the thermodynamic limit is taken. These conditions should be understood as large-system-size statements, and deviations suppressed by sufficiently high powers of system size are allowed. We can construct many more NCFT$_1$ models based on the two given conditions. For example, we can take a $p$-local model and start adding random phases on each qubit (as long as these new  phases are also generated by random uniform fluxes), and we end up with a Hamiltonian that looks like the right-hand side of equation \eqref{eqn:generalProbes}.

We stress that the two conditions we give are sufficient but not necessary for having  NAdS$_2$/NCFT$_1$ physics, as there are regimes that go beyond the $q$-combinatorics of chord diagrams, such as the fixed $p$ large $N$ limit of the SYK model which gives NAdS$_2$/NCFT$_1$ but breaks the second condition. However, the two conditions should not be broken too violently. For example,  we can strongly break the uniformity requirement by assigning an independently random phase to each edge of the hypercube \cite{beckwith2012distribution}, and we end up with radically different combinatorics which will not deliver  NAdS$_2$/NCFT$_1$ physics; a local spin chain model strongly breaks the second condition (much more so than the fixed $p$ SYK), because it can only support nonzero fluxes on a vanishingly small fraction of the Fock-space plaquettes. Hence we would like to suggest that the picture of random uniform Fock-space fluxes is relevant even outside the chord diagram regime, however it is far from clear what the precise characterization is in that case.  Some nontrivial modification is likely needed, since  some  $p$-local models do not reproduce  NAdS$_2$/NCFT$_1$ physics in the fixed $p$ large $N$ limit (even though they do in the double-scaled limit); for example the fixed $p$ SYK model works but the Erd{\H{o}}s-Schr\"oder $p$-spin model likely does not because of a replica-off-diagonal ordering  \cite{baldwin2020}, and a tentative complete characterization must be able to distinguish them. 

\subsection{How can the fluxes arise?}\label{sec:berry}
If random uniform flux is indeed a unifying feature  of NCFT$_1$ microscopic models, we then need an explanation on how such fluxes could conceivably arise in a UV-complete holographic CFT such as the $\mathcal{N}=4$ super Yang-Mills theory. We speculate  that such  disordered fluxes can arise as Berry curvatures \cite{berry1984}. 
Since near-AdS$_2$ is the near-horizon long-throat geometry of a near-extremal black hole,  and as such there is a separation of timescales of the near-horizon and far-away-from-horizon degrees of freedom. Holography suggests  that there should be a separation of timescales on the CFT side as well which entails adiabaticity. We might generically expect Berry curvatures to appear on the slow subsystem once the fast subsystem is integrated out.  The paradigmatic way in which this happens is through the Born-Oppenheimer  approximation (BOA) \cite{mead1979b,mead1980a,bohm1992}, but at the present moment it is unclear to us if the  original BOA works here because our slow degrees of freedom  are chaotic, whereas BOA needs the slow part to be approximately integrable.\footnote{When dealing with the fast degrees of freedom, BOA replaces the slow degrees of freedom by a set of quantum numbers, which are approximately conserved due to the timescale separation. However, this cannot hold if the slow part is chaotic: if there were a set of (approximate) quantum numbers that completely fixes the state, then the state is (approximately) integrable.} It is independently interesting to ask if and how BOA needs to be replaced in such a situation, even in the context of few-body physics.  Since black holes  correspond to chaotic states on the CFT side, the resulting Berry fluxes will be pseudorandom and therefore can be modeled as disorders \cite{Berry_2018, Berry_2019, Penner_2021}, and adiabaticity makes the disorders naturally quenched in time and smoothly varying (though not necessarily strictly uniform) across the slow degrees of freedom. That is, adiabaticity alone already ensures that some parts of the conditions we listed in section \ref{sec:microscopics} are not going to be violated too strongly.  There is no particular reason to think that the fluxes generated this way should be statistically independent,  but this is not a serious obstruction since we know the SYK model with fixed $p$ is a NCFT$_1$ model as well, and its Fock-space flux correlations play a non-negligible role. Finally, Berry phase is a fairly ubiquitous phenomenon in quantum mechanics, which may correspond to the fact that NAdS$_2$ is a fairly ubiquitous near-horizon geometry on the gravitational side.

\section{Operator growth}\label{sec:opGrowth}

Both the Parisi model and the DS-SYK model share the same chord expansion. In this section we suggest that there is a deeper relation which is that a Parisi hypercube model can be used as a fairly universal model for operator growth in $p$-local systems.  Here we will just present the construction and leave its further analysis to future work. We will comment on the relation to the gravitational two-sided Hilbert space, and to Krylov complexity at the end of the section.

\subsection{``Growth'' of the thermofield double with temperature}\label{sec:opGrowthTFD}

The simplest starting point for the construction is to consider the DS-SYK model's thermal density matrix $e^{-\beta H}$, and how it ``evolves" with respect to $\beta$. Recall that $H=\sum_I J_I\Psi_I$, and expand $e^{-\beta H}$ as a sum of products of the $\Psi_I$ (we will discuss  shortly to what extent this is a faithful representation). When discussing the DS-SYK model, we will use the notations of section \ref{sec:importanceOfFluxes}.

Consider now the following hypercube in operator space (i.e., ${\cal H}^\dagger\otimes {\cal H}$). First choose some arbitrary ordering (denoted by $\Lambda$) over the set of multi-indices. 
Denote ordered product over the multi-indices (from smallest to largest) by $\prod_{I\in\Lambda}$. For each vector of $\{0,1\}$, of length $N \choose p$, define the following operators
\begin{equation}
    O({\vec n}) = \prod_{I \in \Lambda} \Psi_I^{n_I},\ \ \ n_I\in \{0,1\}
\end{equation}
so that the expansion above becomes 
\begin{equation}\label{eq:TrmCube}
    e^{-\beta H}=\sum_{\vec n} a_{\vec n} O({\vec n})
\end{equation}
Since $O(\vec n)$ are the points on a hypercube, then we can consider $e^{-\beta H}$ as a state on the hypercube (embedded into states of ${\cal H}^\dagger\otimes {\cal H}$).

Next consider the ``evolution'' of the density matrix with $\beta$, $\partial_\beta e^{-\beta H} = -e^{-\beta H}H$. When written as a wave function on the hypercube above we can consider this evolution as generated by an effective Hamiltonian,  made out of terms that change a single digit in ${\vec n}$, so we exactly have a motion on the hypercube. The main difference from the Parisi model, as we defined it before, is that instead of taking a trace over all the states of the hypercube, we start at the origin ${\vec n}=(0,0,.....)$ which corresponds to $\beta=0$ or the identity operator. The regime in which the computation is strictly valid is when we start from the origin and do any finite number of steps\footnote{Actually, it could be valid also for a number of step $\propto d^\alpha, \alpha<1$.}, i.e., a number of steps which does not scale linearly in $d$. 

The remaining issue is to evaluate the phases. This is straightforward since moving around the plaquette just amounts to a phase
\begin{equation}
\Psi_I\Psi_J\Psi_I\Psi_J=(-1)^{|I\cap J|}
\end{equation}
and the average of the phases is just $\langle (-1)^{|I\cap J|} \rangle=q$. In the SYK model, the phases along different plaquettes have complicated correlations. What the Parisi independent  phase approximation does is replace them by independent random variables on  plaquettes, with the same mean. So the Parisi model is a typified version of the growth of $e^{-\beta H}$ with $\beta$ in the SYK model. In the DS-SYK model, we do not encounter the complicated correlations of the SYK model because we do a finite number of steps [in the large $d=\binom{N}{p}$ limit] around the operator hypercube. This statement is the same as the statement that there are no triple intersections in the DS-SYK limit.

Finally, we would like to comment on the ``faithfulness" of the expansion \eqref{eq:TrmCube}. 
Clearly it is not, since for example there are many relations of the form $\Psi_I\Psi_J=\Psi_{J'}\Psi_{I'}$ for appropriate $I,J,I',J'$, (and even more constraints from higher-order products). But this problem is also solved by taking a finite (but arbitrarily large) number of steps from the origin when $d\rightarrow\infty$. In this limit, such redundancies in the description are of measure zero. If we use the sparse (double-scaled) SYK this problem is further alleviated. This is because the sparse Hamiltonian typically contains $\sim N$ (instead of $N^p$) terms in the random sum, so the system has to time evolve much longer before we run into the redundancies.

The discussion here is a refinement of the discussion in \cite{rabinovici2023bulk} of the Krylov complexity in the DS-SYK model. The authors there start with the so-called $|0\rangle$ state in the chord language, which (for some computational purposes) is the identity matrix on the Hilbert space, i.e., $\mathbb{1}\in {\cal H}^\dagger \otimes {\cal H}$, and the mainly discussed evolution there is the one-sided evolution. The chord number, or Krylov complexity as discussed there, is just the distance $|n|$ from the origin. In section \ref{sec:chordKrylovHamming} we will show that in the hypercube model, this notion of complexity is precisely the coarse description of the microscopic Hamming distance on the hypercube.

\subsection{More general operators}

In section \ref{sec:opGrowthTFD} we considered the evolution of the density matrix of the DS-SYK model as a function of $\beta$. This is not the standard time evolution of a generic operator. So we would like to see whether there is a hypercube growth model for other operators.

It turns out that different classes of operators can have different hypercube growth models. For example, let us consider two classes of operators: a single monomial of fermions $\Psi_I$,  and operators which we will call simple, of the form $O=\sum_{ I} c_{ I} \Psi_{ I}$, where the $c_{I}$ are i.i.d. random numbers. In the SYK model the simple operators are the probe operators used to define correlation functions. The nomenclature might be a bit unusual. Sometimes in literature monomials are referred to as ``simple". As we will see below, they are actually difficult to deal with, whereas what we call simple here have a universal behavior, for example in the chord language we discussed.

\subsubsection{Simple operators}

Let us focus first on simple operators ${\cal O}=\sum c_{\hat I} \Psi_{\hat I}$ for  i.i.d.  Gaussian $c_{\hat I}$, and as before $\Lambda$ is some order on the set of ${\hat I}$. We will denote the order by $\Lambda=\{I_1,I_2,....I_{max}\}$ and the reverse order by ${\bar\Lambda}=\{I_{max},..,I_2,I_1\}$. The relevant hypercube is now 
\begin{equation}
 O({\vec m},{\vec n})= \prod_{\hat J\in \bar\Lambda}\Psi_{\hat J}^{m_{\hat J}} \cdot {\cal O} \cdot 
 {\prod_{\hat I\in\Lambda}}\Psi_{\hat I}^{n_{\hat I}},\ m_{\hat J},n_{\hat I}=0,1.\\
\end{equation}
    Time evolution
\begin{equation}
    \partial_t O(t)= i\left(\sum_I J_I \Psi_I \right) O(t) - i  O(t) \left(\sum_I J_I \Psi_I \right)
\end{equation}
is now simply hopping on two independent lattices, one associated with the left and one with the right.
As before this description is valid only when the number of hopping is finite in the large $d$ limit. 

Finally, there is one interesting subtlety here which complicates things relative to the hypercube model, and that is the fact that states in the hypercube with ${\vec m}+{\vec n}$ fixed are not orthogonal.
For example, suppose we start at some point in the double hypercube which we denote by ${\cal O}$. We then hop by $\Psi_I$ either in the left or in the right hypercube. In this case we have two states that mix and the matrix of norms is
\begin{equation}
    \begin{pmatrix}
        \Tr( {\cal O} \Psi_I \Psi_I {\cal O} ) &  \Tr( {\cal O} \Psi_I {\cal O} \Psi_I  ) \\
        \Tr( {\cal O} \Psi_I {\cal O} \Psi_I ) &  \Tr( \Psi_I {\cal O}  {\cal O} \Psi_I )
    \end{pmatrix}
    \propto 
    \begin{pmatrix}
        1 &  q^\alpha \\
        q^\alpha &   1
    \end{pmatrix}
\end{equation}
where $\alpha$ is determined by the length of ${\cal O}$. This is positive definite with eigenvalues $1\pm q^\alpha >0$. The general structure is similar to the Hilbert space discussed in $q$-Gaussian processes \cite{Speicher1991,Fivel1990, zagier1992,Speicher1993,Speicher1997}, and we expect that the mixing Hilbert space will be nondegenerate in general, but we have not proven this. 

This is also just the inner product for the two-sided Hilbert space discussed in \cite{Lin_2022}. In fact, if we keep track only of the how far we are from the origin in each hypercube, i.e., $|m|$ and $|n|$ separately, then we obtain the gravitational two-sided solution with a particle inserted in the middle with distances $|m|$ and $|n|$ from the left and right \cite{Lin_2022,lin2023symmetry,berkooz2022erepr}.

\subsubsection{Monomials}
We will not discuss much about operators that are single monomials. The complication is that now there is a specific $\pm 1$ factor which occurs when we go from $\Psi_{\hat{I}}\mathcal{O}$ to $\mathcal{O} \Psi_{\hat{I}}$. This means that we should work with a single-sided lattice
\begin{equation}
    O({\vec n})=\mathcal{O} \cdot \prod_{\hat{I} \in \Lambda} \Psi_{\hat{I}}^{n_{\hat{I}}},\ n_{\hat{I}}=0,1
\end{equation}
where the operators $\Psi_{\hat{I}}$ are the ones that appear in the Hamiltonian. Evolution is now hopping on the single lattice but we can hop with one of two phases, depending on whether we apply $\Psi_{\hat{I}}$ on the left or on the right. The regime of validity of this description is as discussed before. 

We will not discuss this further, since it is difficult to make a robust statement due to the phases above.

\subsection{Chord Krylov complexity as coarse-grained Hamming distance}\label{sec:chordKrylovHamming}

We would like to end with a few more words on the notions of Krylov complexity and coarse graining implemented by chords.
It has been shown in the DS-SYK model \cite{Berkooz:2018jqr,Berkooz:2018qkz}  and in the Parisi model itself \cite{Parisi:1994jg}  that the averaged moments   in the above regime can be equivalently computed as the expectation value of a $q$-deformed oscillator system,
\begin{equation}\label{eqn:qOscMoment}
    2^{-d}\vev{\Tr H^k} = \langle 0| \hat T^k | 0\rangle,
\end{equation}
where
\begin{equation}
    \hat T = a+ a^\dagger, \quad, a a^\dagger -q a^\dagger a = 1, \quad a|0\rangle =0.
\end{equation}
The oscillator vacuum state $|0\rangle$ is interpreted as a zero-chord state, and by applying $a^\dagger$ operator  $n$ times on the vacuum state we get an $n$-chord state $|n\rangle$.  Note that the equality \eqref{eqn:qOscMoment} only holds at leading order in $1/d$ and for $k\ll d$, and hence the replacement of $H$ by $\hat T$ is essentially a coarse-graining procedure. In the Parisi model, the correspondence between microscopic operators and the coarse-grained  operators is particularly simple. To see how it works, we replace the trace on the left-hand side of equation \eqref{eqn:qOscMoment} by a state expectation value. The choice for such a state is nonunique,  but for convenience we can choose it to be annihilated by all $T_\mu^-$, namely the position eigenstate $|\vec x\rangle$ with $x_i=-1/2$ for all coordinate components.   It then becomes clear that
\begin{equation}\label{eqn:coarseSubs}
  -\frac{1}{\sqrt{d}}\sum_\mu T_\mu^+ \overset{\text{coarse}}{\longrightarrow}  a^\dagger, \quad   - \frac{1}{\sqrt{d}}\sum_\mu T_\mu^- \overset{\text{coarse}}{\longrightarrow}   a.
\end{equation}
We note that a very similar coarse-graining procedure was proposed in \cite{Speicher1993}. The $n$-open-chord state $|n\rangle$ is then the coarse description of a linear superposition of all microstates that are $n$-Hamming distance (in terms of hypercube coordinates) away from the starting position $\vec{x}$. In \cite{rabinovici2023bulk}, chord states $|n\rangle$ are identified as Krylov basis of the transfer matrix $\hat T$ and the time evolution of a chord state has the form 
\begin{equation}
    |\phi(t)\rangle = e^{-i \hat T t}|\phi(0)\rangle = \sum_n \phi_n(t) |n\rangle,
\end{equation}
which has the chord Krylov complexity 
\begin{equation}
    C(t) = \sum_n  n |\phi_n(t)|^2,
\end{equation}
which is essentially the expectation value of open chord numbers. In the Parisi hypercube model, this quantity has a clear-cut meaning as the coarse description of the average Hamming distance that a microstate has traveled on the hypercubic Fock-space graph.  In DS-SYK, such an identification is less straightforward since the Hamiltonian flips infinite number ($\sim \sqrt{N}$) of qubits.

\section{Subleading moments, sparse SYK, and bulk vertices}\label{sec:subleading}
In \cite{Jia_2020, garcia2021} it was observed and proved that the hypercube model and the (double-scaled) sparse SYK model share the same expression for moments up to subleading order. Equipped with our new understanding that an SYK model is also a model of Fock-space fluxes, we explain why this has to be the case in this section. The gist is that if a model has---on top of the two conditions listed in section \ref{sec:microscopics}---also  an expansion in terms of node degree of its Fock-space graph, then its subleading (in inverse node degree) moment has a unique form independent of the finer details of its microscopics.   This implies the form of the subleading moments is also very ubiquitous, which should be found in many more models other than just the hypercube model and the sparse SYK model. Furthermore, we will observe that when the same subleading consideration is applied to operator probes, we get diagramatics that are suggestive of a bulk interaction vertex with the correct suppression factor due to dimensional reduction down to AdS$_2$. Let us now begin with a very simple rederivation of the subleading combinatorics for the hypercube model, since the approach used for the derivation in \cite{Jia_2020} is unnecessarily complex.
\subsection{Subleading combinatorics for moments}
The general expression for moments $\vev{\Tr H^{2k}}$ is given by equation \eqref{eqn:momentGeneralHopping}, and we noted that the subscripts form $k$ distinct pairs at leading order.  Therefore, such contributions always give a prefactor of $d(d-1)\ldots(d-k+1)/d^k$, multiplying the $q$-moment expressions such as equation \eqref{eqn:HchordRules}.  This implies that the subleading correction to  $2^{-d}\vev{\Tr H^{2k}}$ always has a piece of contribution of the form
\begin{equation}\label{eqn:vestigialSubleading}
- \frac{1}{d} \binom{k}{2} M_{2k}^\text{leading},
\end{equation}
where $M_{2k}^\text{leading}$ is given by equation \eqref{eqn:HchordRules}. 

\begin{figure}
    \centering
    \includegraphics[scale=0.6]{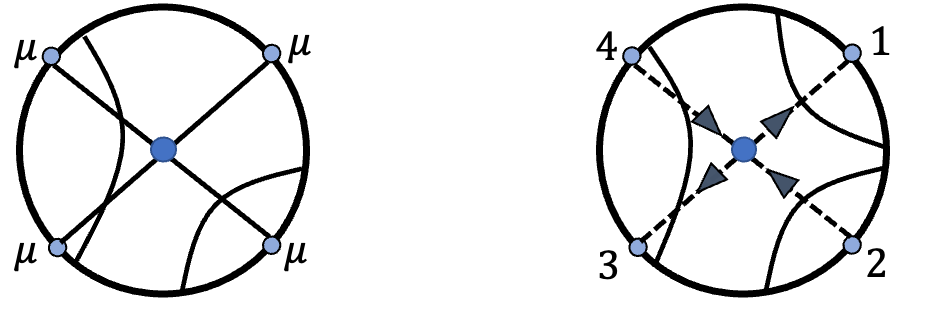}
    \caption{Left: a diagram for a subleading contribution to $\vev{\Tr H^8}$, which evaluates to $q$. Right:
    subleading diagram for probe four-point function; in general the four probes can carry different fluxes, and we draw arrows on the legs to indicate the flux conservation condition \eqref{eqn:fluxConservation4pt}.  This diagram evaluates to $\tilde q^2$ if all four legs carry the same flux. In both figures, the solid circle at the center of the disk means that this is best considered as a contact point (bulk dot) of four legs rather than an intersection of two chords.}
    \label{fig:subleading4pt}
\end{figure}

What is more interesting is the genuinely new subleading contributions that account for the further coincidence of the $k$ pairs of the subscripts. Namely,  hoppings of the form 
\begin{equation}\label{eqn:fourDs}
    \sum_{\mu, \mu_1, \ldots, \mu_{k-2}}\vev{\Tr ( D_{\mu_1}\cdots D_\mu \cdots D_\mu \cdots D_\mu \cdots D_\mu \cdots D_{\mu_{k-2}}\cdots)}
\end{equation}
which mean  that we consider Fock-space paths where one of the directions (the $\mu$ direction) is traversed four times \cite{Parisi:1994jg, marinari1995}.  Hence, this subleading contribution to  $\vev{\Tr H^{2k}}$ is obtained by summing over all possible insertions of a quadruple of identical hoppings into  $\vev{\Tr H^{2k-4}}$. We show one example in the left panel of figure \ref{fig:subleading4pt}, since the quadruple is a group of four operators with identical subscripts, there is no natural way to divide them into two pairs, so we show it as a ``bulk dot" in the disk with four legs reaching out to the circumference.

Upon such an insertion, if an $H$ chord in the $\vev{\Tr H^{2k-4}}$ intersects with odd number of the quadruple legs then we would have exactly one $q$ factor, and if it intersects with an even number of legs we would have none. This is because in equation \eqref{eqn:fourDs}  the hoppings along the $\mu$th direction must occur as 
\begin{equation}
\text{either}   \quad \cdots T_\mu^+ \cdots T_\mu^- \cdots  T_\mu^+ \cdots  T_\mu^- \cdots \quad \text{or} \quad    \cdots T_\mu^- \cdots T_\mu^+ \cdots  T_\mu^- \cdots  T_\mu^+ \cdots.
\end{equation}
So when moving a $D_\nu$ across two $D_\mu$'s, a $T_\nu^\pm$ operator always crosses one  $T_\mu^+$  and one $T_\mu^-$. Then according to the relation \eqref{eqn:VpmAlgebra}, it will get two exactly opposite phases from the two crossings, which cancel. By the same reasoning,  if an $H$ chord crosses three quadruple legs we get a factor of $q$.
Hence, the (genuinely new) subleading chord diagram rule is 
\begin{equation}\label{eqn:trueSubleading}
    M_{2k}^\text{sub} =\frac{1}{d}\sum_{\substack{\text{CD$(k-2)$ with a}\\ \text{ quadruple inserstion}} } q^{ \text{No. of }H-H \text{ inters.}} q^{(\text{No. of } H \text{ chord--quaduple leg inters.}) \text{ mod 2}},
\end{equation}
where the sum is over all chord diagrams with $k-2$ ordinary $H$ chords and a quadruple insertion (so there are $\frac{1}{3}\binom{k}{2}(2k-1)!!$ diagrams). Note that, as expected, the factors are the same regardless of whether an $H$ chord crosses the $D_\mu$ chords to the left or to the right of the bulk dot. 

In  \cite{Jia_2020} the same combinatorics were obtained by considering all possible ways of degenerating lattice paths for $\vev{\Tr H^{2k}}$, and there it was called the ``merge and delete'' prescription: merge refers to merging a pair of chords in $\vev{\Tr H^{2k}}$, which is equivalent to an insertion of the four-legged quadruple into $\vev{\Tr H^{2k-4}}$, and  delete corresponds to the mod 2 operation. The sparse double-scaled SYK model has the same expression \eqref{eqn:trueSubleading} for its subleading moments, which we will explain soon.    The first few moments up to  $M_{18}^\text{sub}$ were already computed explicitly in the early work on the hypercube model \cite{marinari1995}, but a simple algebraic expression for  $M_{2k}^\text{sub}$ is still lacking, except for the special cases of $q=1$ and $q=0$ (see appendix \ref{app:subleadingq1and0}).

\subsection{Probes at subleading order and bulk interactions}\label{sec:subleadingBulkVertex}
Now let us apply the same subleading consideration to probe operators, and we will suggest that these subleading corrections are relevant for bulk interactions. We would like to consider probe hoppings  of the type $\mu\mu\mu\mu$ and relate it a bulk four-point vertex. For probes we have the freedom to assign different fluxes to each probe, namely we can insert
\begin{equation}
    \quad \cdots \tilde D_\mu^{(1)} \cdots \tilde D_\mu^{(2)} \cdots \tilde D_\mu^{(3)} \cdots  \tilde D_\mu^{(4)} \cdots
\end{equation}
where $\tilde D_\mu^{(i)}$ is defined by a flux $\tilde F^{(i)}$.  Again, in terms of forward and backward hoppings, they can only appear in the order of  either 
\begin{equation}
    \cdots \tilde T_\mu^{+,(1)} \cdots \tilde T_\mu^{-,(2)} \cdots  \tilde T_\mu^{+,(3)} \cdots  \tilde T_\mu^{-,(4)} \cdots   
\end{equation}
or
\begin{equation}
     \cdots \tilde T_\mu^{-,(1)} \cdots \tilde T_\mu^{+,(2)} \cdots  \tilde T_\mu^{-,(3)} \cdots  \tilde T_\mu^{+,(4)} \cdots.
\end{equation}
The four fluxes need to obey a constraint, as can be seen by computing
\begin{equation}\label{eqn:mmmmContribution}
\begin{split}
2^{-d}d^{-2}\sum_{\mu}\sum_{x=\pm}\vev{\Tr (  \tilde T^x_{\mu,(1)}  \tilde T^{-x}_{\mu,(2)}  \tilde T^x_{\mu, (3)} \tilde T^{-x}_{\mu,(4)})} = \frac{1}{d} \vev{\cos \frac{\tilde F^{(1)}-\tilde F^{(2)}+\tilde F^{(3)}-\tilde F^{(4)}}{4}}^{d-1}.
\end{split}
\end{equation}
This means we must have
\begin{equation}\label{eqn:fluxConservation4pt}
    \tilde F^{(1)}+\tilde F^{(3)}=\tilde F^{(2)}+\tilde F^{(4)}
\end{equation}
to avoid exponential suppression.  At large $d$ this resembles a four-point vertex that imposes a conservation law for fluxes.  This conservation is naturally satisfied by  the uncrossed correlators  $\vev{O_1 O_1 O_2 O_2}$ and $\vev{O_1 O_2 O_2 O_1}$ considered before.   In the right panel of figure \ref{fig:subleading4pt}, we show this  subleading contribution as a quadruple insertion with four legs (dashed lines), and in this case we attach arrows to the legs to imply the flux conservation.  This mimics a four-point bulk vertex if we think of the disk as an AdS$_2$.
The strength  of this four-point vertex is $1/d$ according to equation \eqref{eqn:mmmmContribution}, and this is consistent with the expectation from compactification of a gravitational theory in the bulk down to two dimensions. For example, consider the following compactification 
\begin{equation}
   \text{ 11D SUGRA} \to \text{AdS}_4\times S^7 \to \text{AdS}_2\times S^2 \times S^7.
\end{equation} 
The first arrow comes from a background solution of supergravity (SUGRA) and the second arrow comes from going to the near-horizon limit of an extremal black hole in AdS$_4$. We anticipate a scalar probe $\varphi$ that descends from one of the components of the metric tensor in 11 dimensions, and let us check what the natural scale for the $\varphi^4$ term would be after compactification. We only need to look at the Einstein-Hilbert term 
\begin{equation}
    S \sim \frac{1}{G_{11}} \int d^{11}x \sqrt{-g} R+\ldots
\end{equation}
where $G_{11}$ is the 11-dimensional Newton's constant. After compactification and taking the lowest modes on $S^7$, we expect AdS$_4$ effective action to be  
\begin{equation}
            S \sim \frac{1}{G_{11}} \int d^{4}x \sqrt{-g}\left[ A (\partial \varphi)^2 +B \varphi^4\right]+\ldots
\end{equation}
where we have only included the kinetic and quartic terms of $\varphi$.  The constants $A$ and $ B$ are powers of the radius $l_{\text{AdS}}$ of $S^7$ and AdS$_4$. Given that $\varphi$ is dimensionless (since it was a metric component),  we deduce that 
\begin{equation}
\begin{split}
S_{\text{AdS}_4} &\sim \frac{1}{G_{11}} \int d^{4}x \sqrt{-g} \left[ l_{\text{AdS}}^7 (\partial \varphi)^2 +l_{\text{AdS}}^5 
\varphi^4\right]+\ldots \\
&\sim  \frac{1}{G_{4}} \int d^{4}x \sqrt{-g} \left[ (\partial \varphi)^2 +l_{\text{AdS}}^{-2}
\varphi^4\right]+\ldots\\
&\sim \int d^{4}x \sqrt{-g}\left[ (\partial \hat \varphi)^2 +(l_{\text{AdS}}^{-2} G_4 )
\hat \varphi^4\right]+\ldots
\end{split}
\end{equation}
where to obtain the second line we absorbed some of the $l_{\text{AdS}}$ dependence into $G_{11}$ to write the action in terms of four-dimensional Newton constant $G_4$, and to canonically normalize the scalar we defined
\begin{equation}
    \hat \varphi := \frac{\varphi}{\sqrt{G_4}}
\end{equation}
which gives us the last line. Now in going to the near-horizon limit of an extremal black hole we encounter a new length scale: the black hole radius $r_0$.  Repeating the same compactification procedure down to AdS$_2$ we get 
\begin{equation}
\begin{split}
S_{\text{AdS}_2} &\sim \int d^{2}x \sqrt{-g}\left[ r_0^2 (\partial \hat \varphi)^2 + \frac{G_4 r_0^2 }{l_{\text{AdS}}^2}
\hat \varphi^4\right]+\ldots \\
&\sim \int d^{2}x \sqrt{-g}\left[ (\partial \bar \varphi)^2  + \frac{G_4 }{r_0^2 l_{\text{AdS}}^2}
\bar \varphi^4\right]+\ldots
\end{split}
\end{equation}
where we obtained the canonically normalized action for the scalar by defining
\begin{equation}
    \bar \varphi := r_0 \hat \varphi = \frac{r_0 \varphi}{\sqrt{G_4}}.
\end{equation}
To obtain the connected four-point function of the boundary from the bulk,  we will need to integrate the bulk vertex of $\bar \varphi$ over the AdS$_2$. If we have a large black hole with $r_0 \gg l_{\text{AdS}}$ then the AdS$_2$ radius will be the same as the AdS$_4$ radius $l_{\text{AdS}}$ \cite{iliesiu2020}, and this cancels the $1/l_{\text{AdS}}^2$ factor.
We then conclude
\begin{equation}
    \vev{\bar \varphi (\tau_1)\bar \varphi (\tau_2)\bar \varphi (\tau_3)\bar \varphi (\tau_4)}_c \sim \frac{G_4}{r_0^2} \sim \frac{1}{S_{\text{BH}}},
\end{equation}
where $S_{\text{BH}}$ is the Bekenstein-Hawking entropy of the black hole. We can similarly consider the reduction of  $\text{10D type IIB} \to AdS_5\times S^5 \to AdS_2\times S^3\times S^5$ and we will end up with the same entropic suppression. Since  $S_{\text{BH}}\sim d$ in the Parisi model, we conclude the subleading suppression factor $1/d$ for a four-point insertion is consistent with the existence of a four-point bulk vertex in the dual AdS$_2$ theory in terms of the interaction strength.  To summarize,  we have two hints which suggest the subleading four-point function of the Parisi model may be dual to a AdS$_2$ four-point vertex:
\begin{itemize}
    \item The corresponding diagram has four legs which cannot be divided into pairs, mimicking a contact interaction. We take figure \ref{fig:subleading4pt} as an inspiration.
    \item  The suppression factor is consistent with the  gravitational compactification result from the bulk. 
\end{itemize}
However, we will need to work out the actual form of the subleading four-point function before we can verify or falsify this identification, which is beyond the scope of the present paper. We can however write down the full chord diagram rule for such four-point insertions, the derivation is entirely analogous to the one that led to equation \eqref{eqn:trueSubleading}. The only complication  is that there can be four different fluxes now, so if an $H$ chord crosses one of the four legs, it gets a factor of 
\begin{equation}
    \tilde q_i = \vev{\cos \frac{F+\tilde F^{(i)}}{2}},
\end{equation}
but if an  $H$ chord crosses two legs, for example $\tilde D_\mu^{(1)}$ and $\tilde D_\mu^{(2)}$, then it gets  a factor of
\begin{equation}
    \vev{\cos \frac{\tilde F^{(1)}-\tilde F^{(2)}}{2}}.
\end{equation}
If all probe fluxes are equal $\tilde F^{(1)}=\tilde F^{(2)}=\tilde F^{(3)}=\tilde F^{(4)}=\tilde F$, then we have again a succinct formula, 
\begin{align}
    &2^{-d} \vev{  \Tr H^{k_4} OH^{k_3} O H^{k_2}O H^{k_1}O}_\text{sub} \\
     &\quad= \frac{1}{d}\sum_{\substack{\text{CD$(k-2)$ with a}\\ \text{ quadruple $O-$insertion}} } q^{\text{No. of } H-H \text{ inters.}} \tilde q^{(\text{No. of } H -O\text{ leg inters.}) \text{ mod 2}}. \nn
\end{align}
If (still a big if) this ``subleading hopping path $\leftrightarrow$ bulk interaction vertex'' correspondence holds, then such Fock-space graph models can be a versatile tool for building AdS$_2$ bulk interactions.  We can imagine using probes that allow  $n$-step  loop paths to model  an $n$-point bulk vertex. For example, we can add to probe operators some hopping terms that hop along the face diagonals of the elementary plaquettes, then there can be nonzero three-point functions.

We do not claim this discussion captures all the subtleties of subleading holography.  For example, \cite{Zhang:2020jhn} discusses the subleading effects of dilatons as an obstruction for finding SYK-like boundary duals. Our discussion does not touch upon such considerations. Moreover, the hypercube model also contains too many light fields, as in the case of various SYK-like models.

\subsection{Connection to the sparse SYK model and the ubiquity of subleading combinatorics}\label{sec:connectionSparseSYK}
As we have seen, the $1/d$ expansion in the hypercube model is essentially a node degree (of the Fock-space graph) expansion: leading order is where every direction is traversed exactly twice,  subleading order is where one of the directions is traversed four times, and subsubleading order will be where one direction is traversed six times or two of the directions are traversed four times, and so on.  Since we have established that the DS-SYK model is in the same class as the hypercube model, we may wish to investigate what happens with its node degree expansion. Analogous to the hypercube model, the subleading (in node degree) term  has the form
\begin{equation}\label{eqn:SYKnodeDegreeExp}
    \Tr(\cdots\Psi_I\cdots \Psi_I \cdots \Psi_I \cdots \Psi_I \cdots)
\end{equation}
where the index set $I$ is summed over. In the SYK model the node degree is $\binom{N}{p}$, so this type of contribution is suppressed by a factor of $N^p$ compared to the leading moment, which is exponentially small in the double-scaled limit where $p\sim \sqrt N$. However, as alluded to in the previous sections, in SYK there are also higher-order corrections from the correlations of the fluxes, which in the $p$-local operator language corresponds to considering the effect of nonzero triple and higher set intersections. It turns out such corrections go as $1/p\sim 1/\sqrt N$ \cite{garcia2018c, Jia_2018, Berkooz:2018jqr}, which is much bigger than the correction from \eqref{eqn:SYKnodeDegreeExp} and makes it invisible in the large $N$ expansion. 

So what does it take to make the SYK type of correction \eqref{eqn:SYKnodeDegreeExp} into a parametrically independent one? One way to achieve it is to simply introduce a second independent parameter into the model that tracks the node degree expansion but is unaffected by flux correlations: the sparse SYK model \cite{swingle2020, garcia2021}  does precisely this.   In the sparse SYK model a second discrete random variable $x_I$ is introduced,
\begin{equation}\label{eqn:sparseH}
    H_{\text{sparse}} = \sum_I  x_I J_I \Psi_I.
\end{equation}
The new random variable $x_I$ takes the value of either $0$ or $1$, and $x_I$ with different subscripts are independently and identically distributed. The expectation value $\vev{x_I}\in [0,1]$ measures how sparse the Hamiltonian matrix is: the smaller $\vev{x_I}$ is, the more sparse the Hamiltonian is. We trade $\vev{x_I}$ for another equivalent parameter $k$ defined by
\begin{equation}
    k N := \vev{x_I} \binom{N}{p}
\end{equation}
where $k$ can have a scaling anywhere between $N^0$ to $N^{p-1}$, and $kN$ is the average number of operator monomials that appear in the right-hand side of \eqref{eqn:sparseH} and thus keeps track of the node degrees of the Fock-space graphs. This second parameter allows us to expand the moments in $1/kN$ while staying in the double-scaled limit. It was observed in \cite{garcia2021} that as a function of the $q$-parameter, such an expansion has the form
\begin{align}\label{eqn:sparseAndParisiRelation}
     (\text{sparse SYK moments})_{\text{double scaled}}= &\text{Parisi leading} +  \frac{3}{kN}\times  \text{Parisi subleading} \nn  \\
     & + O(1/(kN)^2),
\end{align}
where the $1/(kN)^2$ term does not agree with the hypercube model.  One can further study the deviation from the double-scaled result due to flux correlations  of different faces of the Fock-space graph (triple or higher intersections for $I$'s), and the effect is to add  terms suppressed in powers of $1/p$ (or mixed powers of $1/p$ and $1/kN$) to the above expression. In other words, the introduction of the node degree parameter $k$ allows us to perform a double expansion in $1/kN$ and $1/p$ for the sparse SYK, and since $k$ never enters the flux correlation effect, there will be no ambiguity in how to organize the double expansion. 

Hence, the sparse DS-SYK is not only the same as the hypercube model in that they both satisfy the two conditions listed in section \ref{sec:microscopics}, which ensures they have the same leading moments,  but also that they have an expansion with the same graph-theoretic interpretation (the node degree expansion).  This alone does not yet ensure the subleading moments to coincide, after all, the two models are not identical microscopically and we would generically expect higher-order expansion to capture the microscopic differences. After taking the ensemble average, such difference should manifest as different correlation patterns of the fluxes.  We have made sure that fluxes on different plaquettes are independent, but there is always the self-correlation. The self-correlation effects   can be seen when a loop circles the same plaquette for more than once. For example, if a loop circles a plaquette twice, its value is  $\langle\cos 2F\rangle$ which in the hypercube model cannot be simply related to the value of $q=\langle \cos F\rangle$ unless we assume simple forms of the disorder distribution, whereas in the sparse SYK model its value is simply $\vev{\cos 2F}=1$ since $F$ is either 0 or $\pi$.   However, this effect can only be seen from the subsubleading order because to loop around a plaquette twice, two distinct directions must be traversed four times each. 

In short, the reason that equation \eqref{eqn:sparseAndParisiRelation} holds is that we are doing the same expansion on models defined by random uniform and i.i.d. Fock-space fluxes, and although there are differences in the microscopics of the two models, such difference cannot be seen until subsubleading order because flux self-correlations can only be detected by looping a plaquette  more than once. 
With this understanding, it is clear the subleading moments of the hypercube model must also be very ubiquitous, that is, a model  must have a subleading (in node degree) moment in the form of equation \eqref{eqn:trueSubleading}  as long as it satisfies, on top of the two conditions listed in section \ref{sec:microscopics},  also a third condition:
\begin{enumerate}
  \setcounter{enumi}{2}
  \item It has an expansion in large node degree of the Fock-space graph.
\end{enumerate}
 One easy new example to construct would be  to take the double-scaled $p$-spin model \eqref{eqn:ESmodel}, and sparsify it in the same way as the sparse SYK construction, and there are many more. One may have noted that the factor of three in front of the subleading term of equation \eqref{eqn:sparseAndParisiRelation} does not quite follow from this reasoning.  This is an artifact of the Gaussian averaging in the sparse SYK model; namely before Gaussian averaging we have $J_I^4 =(J_I^2)^2$ but after averaging we have $\vev{J_I^4} = 3 \vev{J_I^2}^2$. The ``binary sparse SYK'' model proposed in \cite{tezuka2023} avoids Gaussian averaging and is likely an even closer analog of the Parisi hypercube.


\section{Is there a spectral gap?}\label{sec:spectralGap}
In \cite{Jia_2020}  a connection was made between the Parisi model and the two-site SYK model proposed by Maldacena and Qi \cite{maldacena2018},
\begin{equation}
    H_{\text{MQ}} = H_{\text{SYK}, L}+ H_{\text{SYK}, R} + i\mu \sum_k\psi_L^k \psi_R^k,
\end{equation}
where $\psi_L^k$ and $\psi_R^k$ are two sets of Majorana fermions defined on two spatial sites that satisfy  
\begin{equation}
    \{\psi_a^k, \psi_b^l\} = 2 \delta_{ab} \delta^{kl}, \quad a, b \in \{L, R\}, \quad k, l  \in \{1, \ldots, d\}.
\end{equation}
It was noted that at zero flux the Parisi Hamiltonian $H_0$ (which is also the graph adjacency matrix for the hypercube) is the same as the coupling term of the Maldacena-Qi (MQ) Hamiltonian. Namely when the left and right fermions are written in the appropriate Jordan-Wigner representation \cite{Jia_2020}, we have
\begin{equation}
   \sqrt{d} H_0 = i \sum_k\psi_L^k \psi_R^k.
\end{equation}
We may say the zero-flux Parisi model coincides with the strong coupling limit ($\mu = \infty$) of the MQ model. A natural question follows: if we turn on a small flux on the hypercube, does it retain some MQ physics away from the strong coupling limit?  A useful indicator for the answer is the spectral gap (the gap between the ground state and the first excited state). In the thermodynamic limit of the MQ model, there is an order-one spectral gap  for the full Hamiltonian (in the normalization where the free energy is extensive) induced by the coupling term, which separates the unique ground state from the  continuum states. This gives the MQ model two phases: a traversable wormhole phase below the gap and a two-black-hole phase above the gap. 

It should be noted that the moment method we have used so far is not adequate to determine if there is a unique gapped ground state:  such a state will contribute an exponentially small amount to moments, which is invisible to any order in $1/d$.  In this sense, the gap question is interesting independent of  its possible relation to the MQ model: if a gap remains in the thermodynamic limit, then our previous results derived from the moment method are only valid above the gap scale.
Based on numerical simulations, it was claimed in \cite{Jia_2020} that the Parisi model does possess a spectral gap for flux smaller than $\pi/2$. However, the numerics of \cite{Jia_2020} rely on relatively small values of $d$ (up to $d = 14$, this is the same size as $N=28$ in the SYK). We would like to revisit this question through both a perturbative analysis and a numerical analysis for larger values of $d$ up to $d=27$. The disorder distribution we use will be Parisi's original choice $F_{\mu\nu} = \phi S_{\mu\nu}$ where $ S_{\mu\nu}=\pm 1$ with equal probability, and this is the same choice used in  \cite{Jia_2020}.  We will find the following:
\begin{itemize}
    \item The numerical results conclusively show that there is no gap for $\sqrt{d}H$ for sufficiently large $d$ if the flux $\phi \geq 0.3\pi$. 
    \item The most likely scenario is that $\sqrt{d}H$ (and hence $H$) is gapless for any value of fixed nonzero flux in the large $d$ limit.  
\end{itemize}
 Thus we believe the statements in \cite{Jia_2020} regarding the spectral gap need to be revised: in fact there is no spectral gap for any nonzero flux at large $d$, and the analogy with the MQ model probably does not hold beyond the zero-flux case.  Note that the Hamiltonian used in \cite{Jia_2020} differs by a normalization factor of $\sqrt{d}$ from the one used in this paper,  namely what we write as $\sqrt{d} H$ presently is $H$ in  \cite{Jia_2020}.

As an anticipation of what is to come, in section \ref{sec:H0spectrum} we analytically solve for the eigenvalues and eigenstates for the zero-flux Hamiltonian $\sqrt{d}H_0$, to set up notations for the coming perturbative analysis. In section \ref{sec:perturbation} we  study the  problem perturbatively  and discuss the implications on the gap closing behavior.  In section \ref{sec:numerics} we demonstrate numerically that the gap closes at least for all $\phi\geq 0.3 \pi$, and together with the perturbative analysis this poses strong evidence that the gap should close for all fixed nonzero flux.
Finally, in section \ref{sec:extensiveScalings} we discuss what happens in the scaling regime $\phi \sim 1/\sqrt{d}$; this is the regime where $\sqrt{d}H$ gives rise to a free energy that is extensive in $d$.

\subsection{The spectrum for the zero-flux Hamiltonian}\label{sec:H0spectrum}
From equation \eqref{eqn:symmetricGauge} we get that for zero flux
\begin{equation}
    \sqrt{d}H_0 =  -\sum_{\mu=1}^d \sigma^1_\mu.
\end{equation}
 This is the same as the adjacency matrix for a $d$-dimensional hypercube graph, whose spectrum is well known.  However, we will still solve for the spectrum explicitly here to set up notations that will be useful for the later perturbative calculations.  It is clear that the eigenstates of $H_0$ are built from tensor multiplying the eigenstates of $-\sigma^1$,
\begin{equation}
    |-\rangle = \frac{1}{\sqrt{2}} \begin{pmatrix}
       1\\1
    \end{pmatrix}, \quad    |+\rangle = \frac{1}{\sqrt{2}} \begin{pmatrix}
       -1\\1
    \end{pmatrix}.
\end{equation}
The ground state of $H_0$ is simply
\begin{equation}
    \sqrt{d} H_0|0\rangle :=  \sqrt{d}H_0 |- -\cdots - -\rangle= -d |0\rangle.
\end{equation}
Flipping any one of the $|-\rangle$ to $|+\rangle$ will increase the energy by $2$. We can denote the eigenstates simply by the positions where we have $|+\rangle$, for example 
\begin{equation}
    |2\rangle = | -+--\cdots-\rangle,\quad  |1,3\rangle =| +-+--\cdots-\rangle
\end{equation}
and so on.  The complete set of eigenstates and their energies are 
\begin{equation}\label{eqn:completSetH0}
    \begin{split}
        \sqrt{d} H_0 |0\rangle &= -d|0\rangle\\
        \sqrt{d} H_0 |m_1\rangle &= (-d+2)|m_1\rangle, \quad \\
  &\cdots\\
   \sqrt{d} H_0 |m_1,m_2,\ldots, m_d\rangle &= d|m_1,m_2,\ldots, m_d\rangle,
    \end{split}
\end{equation}
where $\{m_1,m_2,\ldots,m_k\}\subset \{1,2,\ldots, d\}$. Namely the spectrum is 
\begin{equation}\label{eqn:q=1Energies}
    E_k = -d+2k,  \quad k=0,1,\ldots,d
\end{equation}
with degeneracies
\begin{equation}\label{eqn:q=1Degen}
    n_k = \binom{d}{k}.
\end{equation}
The spectrum of $\sqrt{d}H_0$ contains $d$ gaps, each of which is of order $1$. Note the eigenstates have alternating  parity quantum number 
\begin{equation}
    A |m_1,m_2,\ldots,m_k\rangle = (-1)^k |m_1,m_2,\ldots,m_k\rangle,
\end{equation}
where parity operator $A$ was defined in equation \eqref{eqn:parityOper}.

\subsection{A perturbative  analysis}\label{sec:perturbation}
Since the Parisi model lacks a quasiparticle description, nor do we know how to apply Schwinger-Dyson type of technique here, we will simply study perturbation theory using the original degrees of freedom.   Let us treat the small-$\phi$ Hamiltonian as a perturbation of the $\phi = 0$ Hamiltonian $H_0$.  The perturbative analysis in this section serves two purposes: first, it indicates that the ground state mixes strongly with a finite fraction of all levels; second, it will be a benchmark for our numerical analysis in the next section,  in the sense that we will know the numerical results are in a nonperturbative regime when it deviates strongly from the perturbative predictions, and this is where we look for the gap closing (or nonclosing) behavior.

To first order in perturbation theory, the ground state energy is simply
\begin{equation}\label{eqn:firstOrderGroundE}
    \langle 0 |  \sqrt{d}H | 0\rangle = -d \vev{\cos \frac{F}{4}}^{d-1}=-d \cos^{d-1}(\phi/4).
\end{equation}
We can estimate its range of validity by requiring the shift in ground state energy to be smaller than the gap size of $H_0$. Note that the relevant gap size is 4, namely the energy difference between the ground state and the second excited states. This is because the parity quantum number of the first excited states is different from that of the ground state, so their wave functions can never mix. Hence we need
\begin{equation}\label{eqn:groundEfirstOrderPerturb}
 \delta E^{(1)} = \sqrt{d} \langle 0 | H -H_0| 0\rangle =   d (1-\cos^{d-1}(\phi/4)) \ll 4.
\end{equation}
For any fixed value of $\phi$, the above criterion is always violated for sufficiently large $d$. The criterion can be satisfied only if we take a scaling $\phi < 1/d$,  however this regime is of no physical interest to us. The smallest scaling regime that is still physically relevant is $\phi \sim 1/\sqrt{d}$ as we shall see in section \ref{sec:extensiveScalings}. 
This divergence is a familiar phenomenon in many-body systems: a perturbation that is small per degree of freedom can still have a large effect when the total number of degrees of freedom is large. This in itself does not give us hints on whether the gap closes, and for that we need to study the mixing between the unperturbed ground state and the excited states.

How strongly  do other states mix with the ground state? We can consider the second-order perturbative correction to the ground state, which requires the computation of 
\begin{equation}\label{eqn:pertur2ndOrderGen}
   \delta E^{(2)} = - \sum_{|\psi\rangle \neq |0\rangle}\frac{|\langle \psi| \sqrt{d}H |0\rangle|^2}{E_{\psi}-E_0},
\end{equation}
where $|\psi\rangle $ are the eigenstates of $H_0$ excluding the ground state.  in fact, due to parity symmetry, this only receives contributions from $|\psi\rangle  =|m_1,m_2,\ldots,m_{2l}\rangle$ with $1\leq l\leq d/2$.  The first nonzero contribution comes from the second excited states $|\psi\rangle =|m, n\rangle$ with $E_\psi =-d+4$:
\begin{equation}\label{eqn:2mixing}
    \langle m,n| \sqrt{d} H |0\rangle =\sin^2 \left(\frac \phi 4\right)\cos^{d-3} \left(\frac \phi 4\right) \sum_{\mu\neq m,n}S_{\mu m} S_{\mu n} 
\end{equation}
for the Parisi's disorder distribution $F_{\mu\nu} = \phi S_{\mu\nu}$ with $S_{\mu\nu}=\pm 1$. The expression for general disorder is given in appendix \ref{app:degenPerturb}.
Upon taking modulus square and ensemble averaging,  summing over the $|m,n\rangle$ states gives part of the second-order correction \eqref{eqn:pertur2ndOrderGen},
\begin{equation}
    -\frac{1}{4}\binom{d}{2}(d-2)\sin^4 \left(\frac \phi 4\right)\cos^{2d-6} \left(\frac \phi 4\right).
\end{equation}
In fact, the specific choice of the disorder greatly simplifies the perturbative calculation and the mixing of the ground state with any excited state takes the simple form 
\begin{equation}\label{eqn:generalMixing}
    \langle m_1,\ldots,m_{2l}| \sqrt{d}H |0\rangle =(-1)^l \sin^{2l} \left(\frac \phi 4\right)\cos^{d-2l-1}\left(\frac \phi 4\right) \sum_{\mu\neq m_1,\ldots,m_{2l}}S_{\mu m_1}\ldots S_{\mu m_{2l}}. 
\end{equation}
Hence, upon ensemble averaging, the total second-order energy correction to the ground state is
\begin{equation}\label{eqn:secondOrderShift}
\begin{split}
      \delta E^{(2)} = & -\sum_{l=1}^{d/2} \frac{1}{4l} \binom{d}{2l} (d-2l) [ \sin \left(\phi/ 4\right)]^{4l} \ [\cos\left(\phi/ 4\right)]^{2(d-2l-1)} \\
        = &        -[\cos\left(\phi/ 4\right)]^{2d-2}\sum_{l=1}^{d/2} \frac{2l+1}{4l} \binom{d}{2l+1}  \tan^{4l} \left(\frac \phi 4\right).
\end{split}
\end{equation}
We can find a  bound for $| \delta E^{(2)} |$ by simply noting $3l  \geq 2l+1 >2l$:
\begin{equation}
\frac{3}{4}C\geq | \delta E^{(2)} | \geq \frac{1}{2}C,
\end{equation}
where 
\begin{equation}\label{eqn:2ndOrderEsti}
\begin{split}
C =  & [\cos\left(\phi/ 4\right)]^{2d-2}\sum_{l=1}^{d/2}  \binom{d}{2l+1}  \tan^{4l} \left(\frac \phi 4\right)\\
=& -d \cos^{2d-2}(\phi/4) + \frac{1-\cos^{d}(\phi/2)}{2\sin^2(\phi/4)}.
\end{split}
\end{equation}
 An important point the result demonstrates is that the ground state mixes quite generically with all other states with the same parity quantum number, which are half of all states. And although the contribution of a given higher excited state is suppressed by higher powers of $\sin(\phi/4)$, the suppression is offset by  a smaller power in $\cos(\phi/4)$ and a larger degeneracy that is a higher power in $d$, and hence cannot be neglected. Indeed, this second-order correction provides a sizable improvement on the agreement with the numerical results (see figure \ref{fig:perturbBenchmark}). The situation in the MQ model is just the opposite: the ground state only mixes with a vanishingly small fraction of all excited states (see appendix \ref{app:MQmixing} for a proof). This suggests the gap of the Parisi model should eventually close as $d\to \infty$, unlike the MQ model. One may still raise the following objection by simply reversing the logic on its face: suppose you do find a gap by a powerful numerical calculation, then would you not  \textit{post hoc} declare that there must be a hidden structure in the Hamiltonian, and there has to be a smarter way of doing perturbation theory so that the mixing is small? This was exactly the position taken by \cite{Jia_2020}: after observing that a gap seems to remain for all $\phi <\pi/2$ (for small values of $d$), they attempted to find a new basis so that the ground state better approximates the thermofield double state.  Indeed there is some ground to speculate $\phi =\pi/2$ is a special point because this is where the bulk $q$-Hermite density qualitatively changes the shape.  In the following section, we address this numerically by pushing for larger values of $d$  and demonstrate that the  gap closes at least down to $\phi =0.3\pi$, below which there is no more reason to think that some special structure can emerge.  
 
\subsection{A numerical analysis}\label{sec:numerics}
One difficulty with the numerical computation of the Parisi model is that near the spectral edge  the finite-$d$ properties  converge rather slowly to their limits at $d=\infty$, presumably due to the sparse nature of the Hamiltonian. This is contrary to the ordinary SYK model, where for all values of $q$ numerical results for $d$ as small as 14 (same size as $N=28$ SYK Majoranas) no visible spectral gap remains.
For the Parisi model, using the in-built sparse diagonalization algorithm in MATLAB and our cluster resource, we obtain the energies of the lowest two levels for $\phi=0.1\pi, 0.2\pi, 0.3\pi$, and $0.4\pi$ from $d=14$ to $d=27$.  For each value of $d$ and $\phi$ we obtain results for ten disorder realizations.\footnote{In particular, we use ten realizations of magnetic flux for each value of $\phi$ with varying $d$. For example, for $\phi=0.1\pi$ we can generate a flux $F_{\mu\nu}$ for $d=30$, and use subsets of $F_{\mu\nu}$ to generate Hamiltonians from $d=14$ up to $d=27$.  In other words  for each $\phi$ we generate ten disorder realizations instead of $(27-14+1)\times 10 =140$ realizations.}

\begin{figure}
    \centering
    \includegraphics[scale=0.3]{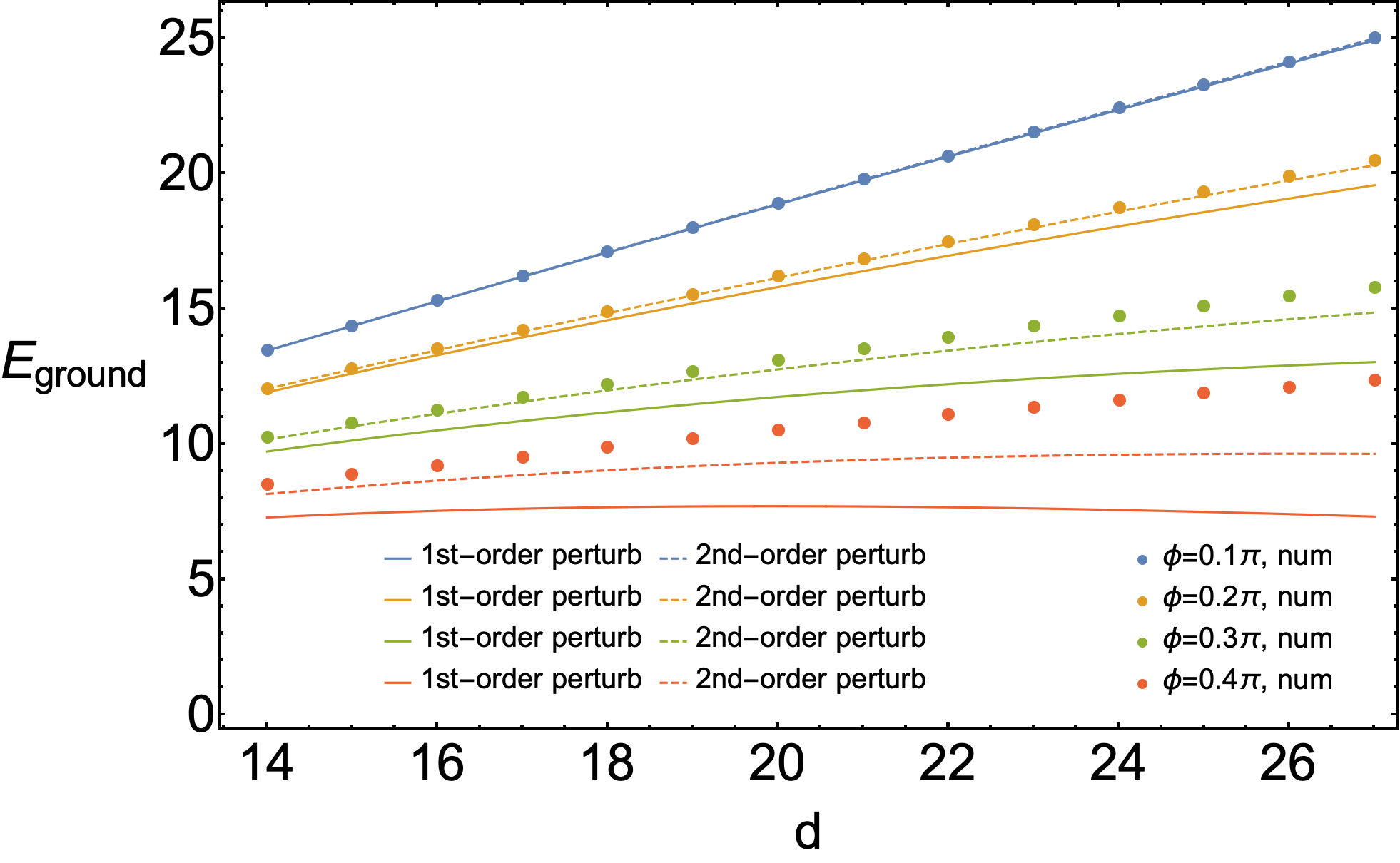}
    \caption{Comparison of the perturbative and the numerical results. For $\phi= 0.1\pi$ the first- and second-order perturbations give nearly identical results, so the solid and dashed blues lines are hard to distinguish. For  $\phi= 0.1\pi$ and $\phi=0.2\pi$ the numerical results agree perfectly with the perturbative calculations, whereas for $\phi= 0.3\pi$ and $\phi=0.4\pi$ the numerical results at larger $d$ show clear deviations from the perturbative results.}
    \label{fig:perturbBenchmark}

    \vspace{0.3cm}
    \includegraphics[scale =0.3]{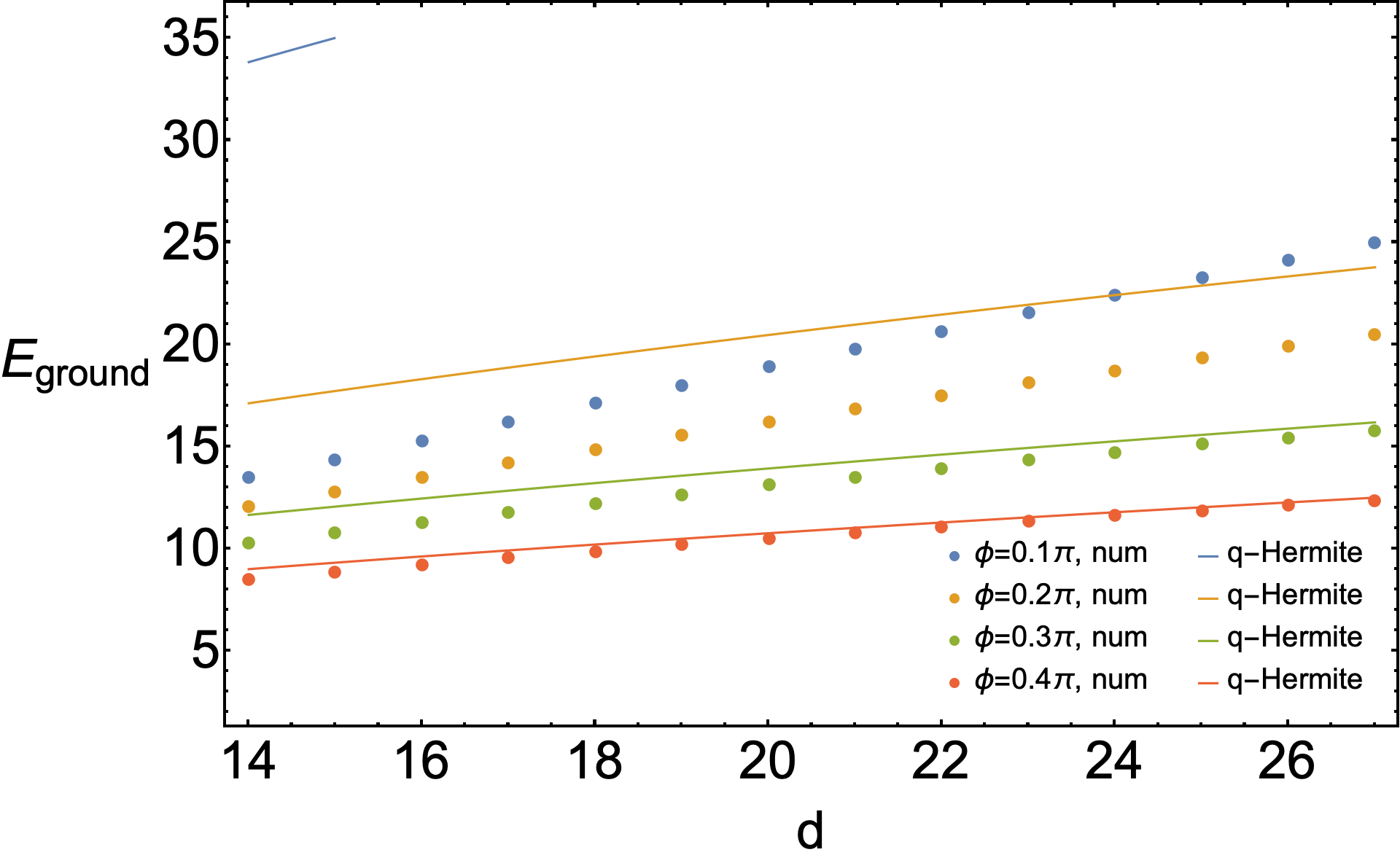}
    \caption{Comparison of the ground state energies with the rescaled $q$-Hermite prediction for the spectral edge [equation \eqref{eqn:rescaledqGaussian}]. The spectral edge is where the continuum spectrum ends and \textit{a priori} does not need to converge to the exact ground state, but we see for $\phi=0.3 \pi$ and $0.4 \pi$ they clearly converge.
    The blue solid line ($q$-Hermite prediction for $\phi=0.1 \pi$) is truncated at the top of the figure since it is well above other scales of the plot.}
    \label{fig:groundQHcompare}
\end{figure}

In figure \ref{fig:perturbBenchmark} we compare (the absolute values of) the numerical ground state energies with the perturbative results.
 As we can see, up to $d=27$,  perturbative results are still very accurate for $\phi=0.1\pi$ and $\phi=0.2\pi$, so we cannot tell if their gaps close only based on the numerical results. On the other hand,  for $\phi=0.3\pi$ and $0.4\pi$ the numerical results clearly diverge from the  perturbative results for larger values of $d$, and we should look at the gap behavior in these nonperturbative regions. For these two fluxes, we present two pieces of evidence that their gaps close. First is the comparison with the spectral edge---namely the end of the continuous part of the spectrum---predicted by the $q$-Hermite density,\footnote{The moment method predicts that spectrum of $H$ is a  $q$-Hermite density whose edges are at $ \pm {2}/{\sqrt{1-q}}$.}
 \begin{equation}\label{eqn:rescaledqGaussian}
  E_{\text{edge}} = \pm \frac{2\sqrt{d}}{\sqrt{1-q}}=\pm \frac{2\sqrt{d}}{\sqrt{1-\cos \phi}}.
\end{equation}
We plot the comparison in figure \ref{fig:groundQHcompare} and see that the ground state energies of $\phi=0.3\pi$ and $0.4\pi$ well converge to the continuum spectral edge \eqref{eqn:rescaledqGaussian}, implying the closure of the gaps. In fact, by comparing figures \ref{fig:perturbBenchmark} and  \ref{fig:groundQHcompare}, we see the results converge to the continuum spectral edges as soon as they start deviating significantly from the perturbative predictions.
\begin{figure}
    \centering
    \includegraphics[scale=0.2]{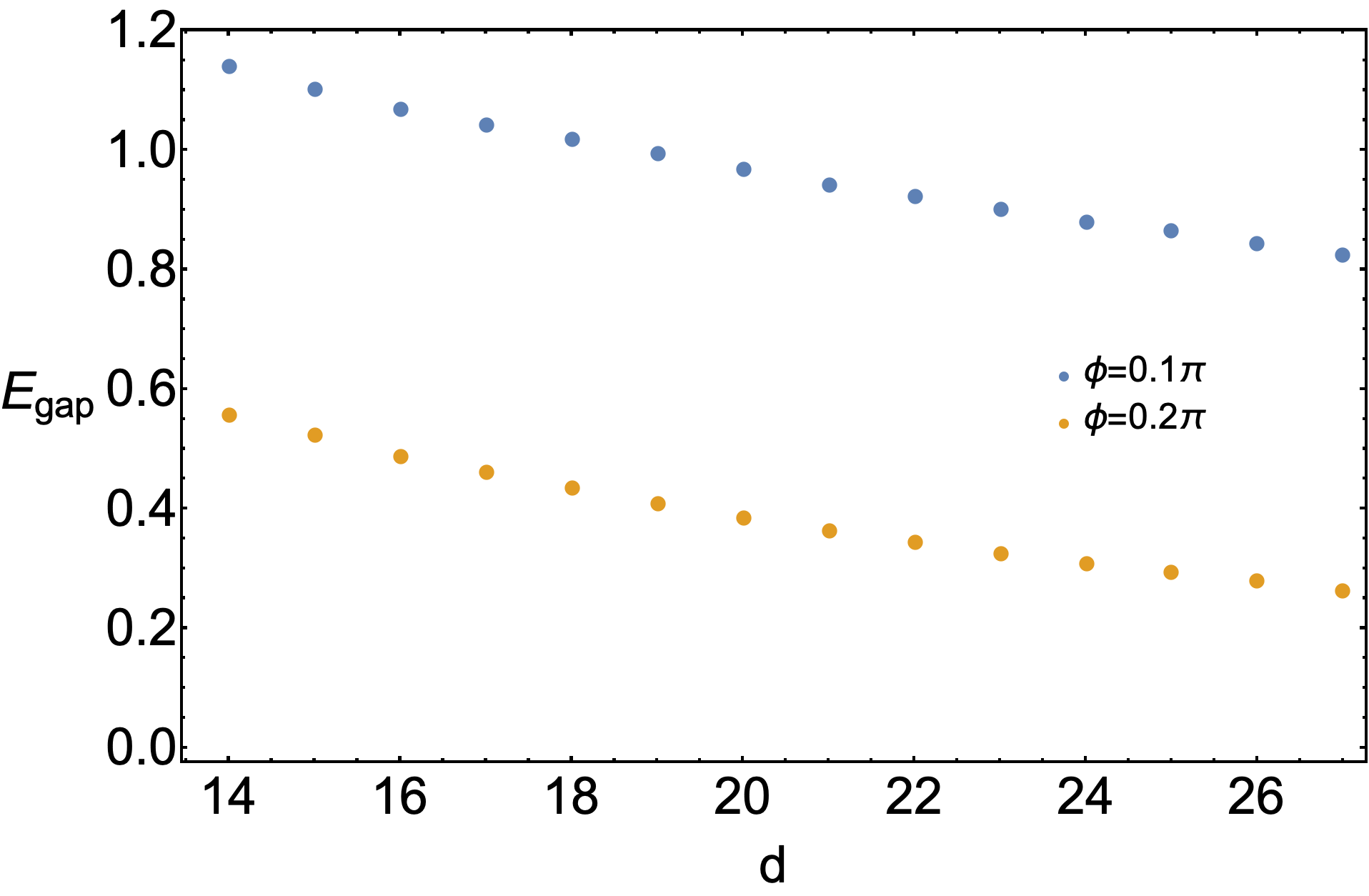}
        \includegraphics[scale=0.2]{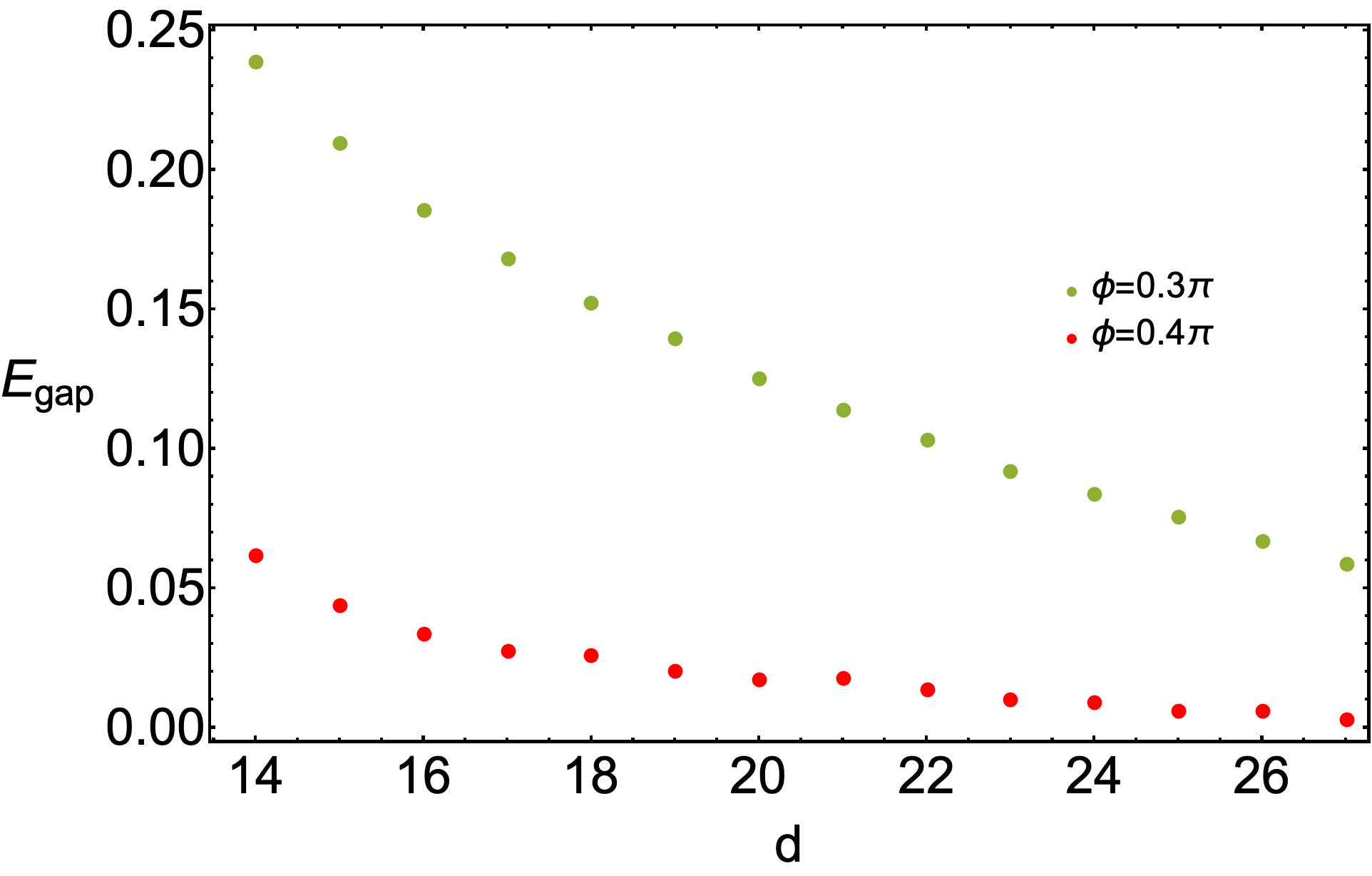}
    \caption{Ground-state gaps as functions of $d$. Left: $\phi=0.1\pi$ and $\phi=0.2\pi$. Right: $\phi=0.3\pi$ and $\phi=0.4\pi$.  Gaps quite clearly close for $\phi=0.3\pi$ and $\phi=0.4\pi$. }
    \label{fig:gapPlot}
\end{figure}
Second,  we plot the numerical spectral gaps as functions of $d$ in figure \ref{fig:gapPlot},  and it is very clear that the $\phi= 0.3\pi$ and $\phi=0.4\pi$ gaps close for large $d$. At $d=27$, the gaps' ensemble fluctuations (not plotted) are comparable to the average gap sizes. Especially for $\phi=0.4\pi$, the gaps become so small as early as $d=17$ that the effects of disorder fluctuations start making the curve look ragged rather than smooth.

Strictly from the numerical evidences, we can only conclude that the gaps close for $\phi \geq 0.3\pi$.  However, from the previous perturbative discussions we expect gaps to close generically for any $\phi \neq 0$ due to strong mixing,  unless some special hidden structure emerges for the Hamiltonian. It was suggested by \cite{Jia_2020} that $\phi =0.5\pi$ could be such a special point, but the numerical results in this section quite conclusively rule out this possibility, and there is no good reason to believe any other smaller flux is special.  Therefore, we believe there is no spectral gap that survives the $d\to\infty$ limit at any fixed nonzero flux.

\subsection{The extensive scaling}\label{sec:extensiveScalings}
 The scalings we have used so far ($H$ or $\sqrt{d} H$) do not give extensive free energy (except for the zero-flux Hamiltonian  $\sqrt{d}H_0$). in fact, any overall scaling of $H$ cannot help because the second cumulant and the fourth cumulant cannot be made simultaneously extensive in $d$, which means we must scale $q$ as well.   We suggest $1-q \sim \alpha/d$ (or $\phi \sim \sqrt{2\alpha/d}$) is such a scaling for $ \sqrt{d}H$, where $\alpha$ is an order-one positive constant.  This is analogous to how one formally obtains the large $p$ SYK from DS-SYK. 
There are two pieces of evidence for the extensiveness for this scaling. First, we can work out the high-temperature expansion of the free energy (say, using the exact results for moments listed in \cite{marinari1995}, up to the eighth moment), and it is extensive in $d$.\footnote{Strictly speaking we only have the results of some low-order cumulants which are extensive in $d$, and we have not proved the extensiveness to all orders of the high-temperature expansion.} 
Second, in such a scaling the ground state energy seems to be extensive in $d$. For example, taken at face value the first-order perturbative result \eqref{eqn:firstOrderGroundE} predicts that the ground state energy moves to 
\begin{equation}
    -d \cos^{d-1}(\phi/4) \to -d e^{-\frac{\alpha}{16}},\quad  \text{as $\cos\phi \to 1-\alpha/d$.}
\end{equation}
and the second-order perturbative correction \eqref{eqn:secondOrderShift}--\eqref{eqn:2ndOrderEsti} is extensive in $d$ as well in this scaling.  Furthermore, the $q$-Hermite rescaling gives 
\begin{equation}
    -\frac{2\sqrt{d}}{\sqrt{1-\cos\phi}} \to -\frac{2d}{\sqrt{\alpha}},\quad  \text{as $\cos\phi \to 1-\alpha/d$.}
\end{equation}
Both estimates are consistent with an extensive ground state energy.
However, we caution that we should not take the exact values of these two estimates too seriously. For perturbation theory, the reason is obvious;  for  the $q$-Hermite rescaling, other than the fact that this scaling is not strictly within the range of validity of the moment method we used, there is also a hard failure because it can violate the diamagnetic inequality: if $\alpha<4$, this result would be smaller than $-d$ which is the ground state energy of $H_0$ without any magnetic field.  

The point on  the diamagnetic inequality might require some explanation. The inequality states that an external magnetic field cannot decrease the ground state energy of a spinless particle.  The inequality is normally phrased for the covariant derivatives $-i \vec{\nabla}-\vec{A}$ in the continuum; here we give a simple proof of it for the Parisi model based on moments.  We consider the situation of finite $d$ and before any ensemble averaging.  It is clear that
\begin{equation}\label{eqn:momentInequality}
    \Tr (H^{ 2l} )  \leq \Tr (H_0^{2l})
\end{equation}
because the contribution of each Wilson loop becomes smaller in the presence of a magnetic flux, namely each Wilson loop gives 1 for $ \Tr (H_0^{2l})$ and gives a cosine for $\Tr (H^{2l})$. We in turn have 
\begin{equation}
    \Tr(e^{-\beta H}) \leq  \Tr(e^{-\beta H_0}) 
\end{equation}
by simply Taylor expanding the exponentials and applying the inequality \eqref{eqn:momentInequality}.  Taking $\beta\to \infty$ picks out the ground  states on both sides, and we conclude 
\begin{equation}
    \langle \Omega |\sqrt{d}H |\Omega\rangle \geq     \langle 0 |\sqrt{d}H_0 |0\rangle =-d.
\end{equation}
 Since the Parisi model has a sublattice symmetry $\Gamma_5$ so that eigenvalues always come in $\pm$ pairs, this also implies that the largest positive eigenvalue of $\sqrt{d}H$ is smaller than $d$. We can also infer from this that the annealed free energy density of the model is not divergent at very low temperatures, unlike that of the fixed $p$  Erd{\H{o}}s-Schr\"oder model  \cite{baldwin2020}. 
 
 Can we say anything about the gap behaviors in this scaling?  The perturbative results \eqref{eqn:secondOrderShift}--\eqref{eqn:2ndOrderEsti} suggest that the mixing of states is still strong in this scaling, thus we are inclined to say the gap should also close, but to end on a cautious note we would like a second method to corroborate this,  which we do not have at the moment.

\section{Conclusion and outlook}
Using chord diagram techniques we demonstrate that the Parisi  hypercube model has the same correlation functions as those of the double-scaled SYK model. In addition, this model exhibits random matrix universality in its energy level correlations as was shown in a previous work \cite{Jia_2020}. Hence, the hypercube model is an equally good microscopic construction of NAdS$_2$/NCFT$_1$ near-horizon holography. We further demonstrate that it is unlikely that the model has any spectral gap as long as the flux is nonzero, thereby revising the claim made in \cite{Jia_2020}. In our view the main value of this model is that its definition looks significantly different from the SYK model and other $p$-local models. This stark contrast serves as a good filter for us to discern which aspects of the microscopics are essential and which are spurious for NAdS$_2$/NCFT$_1$ physics. Our conclusion is that  having a large amount of random uniform fluxes in the Fock space is the important feature for NCFT$_1$ microscopics. This characterization is made precise for all the aforementioned models in the regime  solvable by chord diagrams.  This has several implications. First, it enlarges our toolbox for model building, since (double-scaled) $p$-local models are a proper subset of the class of models we characterized. More specifically, $p$-local operators play two roles at the same time: they provide a Fock-space graph structure and  generate a lot of Fock-space fluxes by virtue of their noncommutativity and nonlocality. But this is not necessary and these two roles can be played by separate objects as we saw from the hypercube model.  Second, this may mean when we are looking for an NCFT$_1$ by RG flowing a higher-dimensional holographic CFT, we should more broadly look for signatures of such Fock-space fluxes, and a speculative possibility is that such fluxes could arise as Berry curvatures. 

Finally, let us end the discussion by making a list of some   puzzles we would like to see solved  and some unexplored applications of the current work, some of which have already been stated in the main text: 
\begin{itemize}
    \item The precise version of the characterization of NCFT$_1$ microscopics we gave in section \ref{sec:microscopics} only applies to models solvable by $q$-combinatorics of chord diagrams. However, the $p$-body SYK model with a fixed $p$ is also a valid NCFT$_1$ model but does not strictly satisfy this characterization. We would like to find a way to expand our current characterization to include such models as well. Perhaps a useful case study is the $p$-spin model studied by Erd{\H{o}}s and Schr\"oder \cite{erdos2014}.  It was pointed out in \cite{baldwin2020} that the fixed $p$ version of this model develops some kind of ordering at low energy, which is also consistent with the level statistics study of a very similar model \cite{hanada2023model}, and thus is unlikely to be a good model for NCFT$_1$. Yet its double-scaled limit has identical behavior to the double-scaled SYK.  It maybe worthwhile to understand this contrast in terms of microscopics. 
    \item Related to the first point, it would be useful to have a better mathematical control of the hypercube model with the scaling $\langle F^2 \rangle \sim 1/d$, which we briefly discussed in section \ref{sec:extensiveScalings}. By analogy, this scaling could behave like the first large $N$ then large $p$ limit of the SYK model.  This could be an entry point for understanding the more conventional limits of the SYK model in terms of Fock-space fluxes. Yet another related question is if Schwinger-Dyson or $G\Sigma$ action techniques are applicable in any limit of the hypercube model.
    \item We speculated that a timescale separation and Berry curvature could be the dynamical origin for the Fock-space fluxes we need. To test the feasibility of this idea, it is desirable to find a toy model with slow and fast degrees of freedom,  in which if the fast ones are integrated out we could end up with a lot of random fluxes.  in fact, it is not entirely clear to us what this ``integrate out'' procedure should look like because the slow degrees of freedom we are interested in are chaotic, as was explained in section \ref{sec:berry}. A clearer understanding of this even in the case of few-body physics is very welcomed. 
    \item We pointed out in section \ref{sec:subleading} that the form of the subleading moments of the hypercube model is also ubiquitous and more than a mere coincidence between the hypercube model and the sparse SYK model. It is therefore interesting to find a solution to the subleading moment problem.  A related fact is that at subleading order the insertion of four probe operators shows a qualitatively correct behavior of a nontrivial conformal four-point function, which could be interpreted to be from a bulk four-point interaction vertex. We would like to see an explicit solution of this as well.
    \item It is clear we can incorporate charge structures using the hypercube hopping operators, by noting that $[\sum_\nu \sigma^3_\nu/2, T_\mu^{\pm}] = \pm T_\mu^{\pm}$. We may try to use this fact to construct NCFT$_1$ models  with conserved $U(1)$ charges and/or supersymmetries,  analogous to those  SYK-type models such as the complex SYK and supersymmetric SYK. It should be explored what new possibilities this new construction could offer.
\end{itemize}

 \acknowledgments{We would like to thank Alexander Abanov, Pawel Caputa, Antonio Garc\'\i a-Garc\'\i a,  Pratik Nandy, Dario Rosa, Ruth Shir,  Erez Urbach, Jacobus Verbaarschot and Zhuoyu Xian for illuminating discussions, and  Rohit Kalloor, Shai Chester and Revital Ackler for their help with the WEXAC HPC cluster. This work is supported by the ISF (2159/22), PBC (76552401) and by the DIP foundation. YJ is additionally supported by the United States–Israel Binational Science Foundation (BSF) under Grant No. 2018068, by the Minerva Foundation with funding from the Federal German Ministry for Education and Research, by the Koshland postdoctoral fellowship and by a research grant from Martin Eisenstein.}

 \pagebreak
\appendix

\section{Rotations in the symmetric gauge}\label{app:rotations}
To see why the Hamiltonian \eqref{eqn:symmetricGauge} is in a symmetric gauge,  let us check how the Hamiltonian transforms under rotations.  Consider a counterclockwise $\pi/2$ rotation in the $\kappa-\lambda$ plane ($\kappa<\lambda$) about the center of the hypercube, namely a rotation implemented by 
\begin{equation}\label{eqn:rotation}
  \begin{pmatrix}
     x'_\kappa \\x'_\lambda 
    \end{pmatrix}=  \begin{pmatrix}
     0 & -1\\
     1 & 0
    \end{pmatrix}\begin{pmatrix}
     x_\kappa\\x_\lambda 
    \end{pmatrix}
\end{equation}
 in the $\kappa$ and $\lambda$ indices and identity in all the remaining $d-2$ indices. 
Using the  qubit representation \eqref{eqn:qubitBasis}  of the lattice positions,  we find the unique transformation that implements the rotation in the tensor product representation:
\begin{equation}\label{eqn:rotTensor}
    R_{\kappa\lambda} := \sigma^1_\lambda P_{\kappa\lambda},
\end{equation}
where $\sigma^1_\lambda$ is the first Pauli matrix acting on the $\lambda$th qubit subspace and $P_{\kappa\lambda}$ simply permutes states in the $\kappa$th subspace and the $\lambda$th subspace.
In a gauge-fixed system a purely geometric transformation like $ R_{\kappa\lambda}$ normally would not respect the gauge-fixing condition,  and an extra gauge transformation is needed.  This is the case if we used the axial gauge,  but for the symmetric gauge \eqref{eqn:symmetricGauge} $ R_{\kappa\lambda}$ is enough.  Let us now demonstrate this by rotating the physical flux and write the Hamiltonian using the gauge \eqref{eqn:symmetricGauge} with the rotated flux.
Under the rotation \eqref{eqn:rotation} the field strength transforms as
\begin{align}
    & F_{\kappa \lambda} \to F_{\kappa \lambda}' = F_{\kappa \lambda},\\
    &F_{\mu\nu} \to F_{\mu\nu}' = F_{\mu\nu} \quad \text{for }\mu,\nu \neq \kappa,\lambda, \\
     &F_{\kappa\nu} \to F_{\kappa\nu}' = F_{\lambda\nu} \quad \text{for }\nu \neq \lambda, \\
    &F_{\lambda\nu} \to F_{\lambda\nu}' = -F_{\kappa\nu} \quad \text{for }\nu \neq \kappa.
\end{align}
In the symmetric gauge, the parallel transporters with the transformed field strength are
\begin{equation}
  T^{+}_\mu{}' = \prod_{\nu\not=\mu} e^{\frac i4 F_{\mu\nu}'\sigma^3_\nu} \sigma^+_\mu.
\end{equation}
A simple calculation would confirm  that 
 \begin{align}
      T^{+}_\mu{}' &= R_{\kappa\lambda } T^{+}_\mu \,  R_{\kappa\lambda }^{-1}, \quad \mu \neq \kappa,\lambda, \\
       T^{+}_\kappa{}' &= R_{\kappa\lambda } T^{+}_\lambda \, R_{\kappa\lambda }^{-1}, \\
       T^{+}_\lambda{}' &= R_{\kappa\lambda } T^{-}_\kappa \, R_{\kappa\lambda }^{-1}.
\end{align}
Hence the transformation rules on $ D_\mu =T^{+}_\mu+T^{-}_\mu $ are simply 
\begin{align}
     D_\mu' &= R_{\kappa\lambda }  D_\mu R_{\kappa\lambda }^{-1}, \quad \mu \neq \kappa,\lambda, \nn \\
       D_\kappa' &= R_{\kappa\lambda } D_\lambda R_{\kappa\lambda }^{-1}, \\
       D_\lambda' &= R_{\kappa\lambda }  D_\kappa  R_{\kappa\lambda }^{-1}, \nn
\end{align}
and finally 
\begin{equation}
    H' = R_{\kappa\lambda } H R_{\kappa\lambda }^{-1}.
\end{equation}
 Note that a rotation is not a symmetry before taking the ensemble average. We stress again that $ D_\mu'$ and $H'$ are in the same gauge as $D_\mu$ and $H$.  Hence a rotation on parallel transports is implemented by a flux-independent similarity transformation. This is quite different from the situation in the axial gauge used by Parisi, where after a rotation a compensating gauge transformation must be implemented so that the gauge-fixing condition is respected \cite{Jia_2020}. It is in this sense that the new gauge we are using is a  symmetric one. 

 \section{Some results in the axial gauge }\label{app:axialGaugeHami}
 For convenience let us copy  definition \eqref{eqn:linkOriginalDef} of link variables in the axial gauge here,
 \begin{equation*}
U_\mu (\vec x) = e^{i \sum_{\nu=1}^{\mu-1} F_{\mu \nu}x_\nu }, 
\end{equation*}
Note in Parisi's original convention $x_\nu =0$ or $1$, but here we follow the convention in the main body of our paper and let $x_\nu = \pm 1/2$.
 We note that  the $\mu$ index on the  right-hand side of the definition must take a positive value. Strictly speaking this definition is only meant for links pointing toward positive directions  and the links pointing toward negative directions are defined as inverses of equation \eqref{eqn:linkOriginalDef}.  Note that since we are on a hypercube, once the vertex position $\vec{x}$ and link axis $\mu$ are specified,  it is entirely determined whether the link is pointing in the positive direction $\hat{e}_\mu$ or negative direction $-\hat{e}_\mu$: the link points positive if $x_\mu=-1/2$ and negative if $x_\mu=1/2$.  We utilize this fact and rewrite equation \eqref{eqn:linkOriginalDef} to be valid for any $\mu$ and $\vec{x}$:
\begin{equation}\label{eqn:linkGeneralDef}
U_\mu (\vec x)= e^{-i\text{sgn}(x_\mu) \sum_{\nu=1}^{\mu-1} F_{\mu \nu}x_\nu }= e^{-i2x_\mu \sum_{\nu=1}^{\mu-1} F_{\mu \nu}x_\nu },
\end{equation}
where $\mu$ is still positive,  but $\mu$ only labels the link axis not the link direction. If we consider a Wilson loop  $\mathcal{C}$ of an elementary plaquette for which the starting point is $\vec{x}$,  the first step is along the $\mu$ axis and the second step is along the $\nu$ axis,  the value of the loop is
\begin{equation}\label{eqn:wilsonLoop}
W(\mathcal{C}) = U_\mu (\vec x)U_\nu(\vec{x} +\hat{e}_\mu)U^{-1}_\mu(\vec{x} +\hat{e}_\nu)U^{-1}_\nu(\vec{x})  =e^{-i\text{sgn}(x_\mu x_\nu)F_{\mu \nu} },
\end{equation}
where $\vec{x} +\hat{e}_\mu$ and $\vec{x} +\hat{e}_\nu$ are interpreted as mod 2 sums. In other words,
\begin{equation}\label{eqn:tranlationGeneralDef}
\vec{x} +\hat{e}_\mu = (x_1, x_2, \ldots, -x_\mu,  x_{\mu+1},\ldots,x_d)
\end{equation} 
With the above notations we can write down the matrix elements of the Hamiltonian simply as
\begin{equation}\label{eqn:hamiAxialGaugeCompact}
H_{\vec x, \vec y} =  -\frac{1}{\sqrt{d}}\sum_\mu  U_{\mu}(\vec x) \delta_{\vec x +\hat{e}_\mu,\vec y}.
\end{equation}

Analogous to the symmetric gauge situation, we may consider operators $O$ in the same statistical class as the axial gauge Hamiltonian, that is
\begin{equation}\label{eqn:axialprobe}
(O_\text{axial})_{\vec x, \vec y} = -\frac{1}{\sqrt{d}} \sum_\mu  \tilde U_{\mu}(\vec x)\delta_{\vec x +\hat{e}_\mu,\vec y},
\end{equation}
where again $\vec x +\hat{e}_\mu$ is  a mod 2 sum as defined in equation \eqref{eqn:tranlationGeneralDef}, and
\begin{equation}
\tilde U_\mu (\vec x)= e^{-i\text{sgn}(x_\mu) \sum_{\nu=1}^{\mu-1} \tilde F_{\mu \nu}x_\nu }.
\end{equation}
where $\tilde F_{\mu\nu}$ may or may not correlate with $F_{\mu\nu}$.
An important fact is that although $H_\text{axial}$ is gauge equivalent to $H_\text{symmetric}$  and $O_\text{axial}$ is gauge equivalent to $O_\text{symmetric}$,  correlation functions of $O_\text{axial}$ are not the same as $O_\text{symmetric}$. This is  because the gauge transformation that transforms    $H_\text{axial}$ to  $H_\text{symmetric}$ is not the same as the one that transforms $O_\text{axial}$ to  $O_\text{symmetric}$. In other words,
\begin{equation}
\begin{split}
        M_1 H_\text{symmetric} M_1^{-1} = H_\text{axial},\\
        M_2 O_\text{symmetric} M_2^{-1} = O_\text{axial},\\
\end{split}
\end{equation}
but 
\begin{equation}
    M_1 \neq M_2.
\end{equation}
Alternatively, one can define probe operators in the axial gauge as 
\begin{equation}
    O'_\text{axial}:=  M_1 O_\text{symmetric} M_1^{-1},
\end{equation}
which would give the same results for correlators as the symmetric gauge,  however, $O'_\text{axial}$ will look complicated in the axial gauge.  Let us study the correlators of $O_\text{axial}$ defined in \eqref{eqn:axialprobe}, and we will omit the subscript ``axial'' from now on.  To compute the one-point moment, we first note
\begin{equation}
2^{-d}\vev{\Tr(- \tilde D_\mu H)} =\frac{1}{\sqrt{d}}\vev{\cos \left(\frac{F_{\mu\nu}-\tilde F_{\mu\nu}}{2}\right)}^{\mu-1}.
\end{equation}
So we arrived at a somewhat peculiar result:  the contraction is $1/\sqrt{d}$ suppressed for $\mu = O(1)$, but exponentially suppressed for $\mu \gg 1$ (say $\mu\sim d$). This reflects the fact that the axial gauge is highly asymmetric.  Hence
\begin{equation}
\begin{split}
2^{-d}\vev{\Tr(H O)} = \frac{1}{d}\frac{1-\vev{\cos \left(\frac{F_{\mu\nu}-\tilde F_{\mu\nu}}{2}\right)}^d}{1-\vev{\cos \left(\frac{F_{\mu\nu}-\tilde F_{\mu\nu}}{2}\right)}},
\end{split}
\end{equation}
which is $1/d$ (instead of exponentially) suppressed.  These results for suppression hold for general one-point insertions  $\vev{\Tr(H^m O)}$.  

The finite correlators start from two-point insertions $2^{-d}\langle\Tr (H^{2m-k-2} O H^k O)\rangle$. As we have shown, to leading order there is no contraction between any $H$ and $\tilde D_\mu$. 
 Intersections among $H$ chords just give powers of $\vev{\cos F}$,  and for intersections between the $O$ chord  and multiple  $H$ chords  we will need to compute the mixed holonomy 
 \begin{equation}\label{eqn:modifiedPlaquetteSameDisorder}
\tilde U_\mu(x) U_\nu(x+\hat e_\mu) \tilde U^{-1}_\mu(x+\hat e_\nu) U^{-1}_\nu(x) = \begin{cases} e^{-i{ \text{sgn} (x_\mu  x_\nu) \tilde F_{\mu\nu}}}\quad \text{if $\nu<\mu$,}\\
 e^{-i{ \text{sgn} (x_\mu  x_\nu) F_{\mu\nu}}}\quad \text{if $\nu>\mu$.}
\end{cases}
\end{equation}
Then for each interlacing ordering $\tilde D_\mu H \tilde D_\mu H$ we would get
\begin{equation}\label{eqn:chordDmu}
\frac{\mu-1}{d} \langle \cos\tilde F\rangle +\frac{d-\mu}{d} \langle\cos F \rangle,
\end{equation}
Again, the $\mu$ dependence is a reflection of the highly asymmetric nature of the axial gauge.  The total contribution from  $H-O$ chord intersections is
\begin{equation}
       \frac{1}{d} \sum_{\mu=1}^d \left[\frac{\mu-1}{d} \langle \cos\tilde F\rangle +\frac{d-\mu}{d} \langle\cos F \rangle  \right]^{\text{\text{No. of }$H$-$O$ int.}} = \frac{\int_{\langle\cos{F}\rangle}^{\langle\cos{\tilde F}\rangle} \tilde u^{\text{\text{No. of }$H$-$O$ int.}} d\tilde u}{\langle\cos\tilde{F}\rangle-\langle\cos{F}\rangle}  + O(d^{-1}).
       \end{equation}
So $O$ behaves like a uniform statistical mixture of all possible $\tilde D_\mu$.

 \section{Algebra and motion on the hypercube}\label{app:EOM}
We would like to take advantage of the simplicity of the hypercube geometry and try to gain some elementary intuition of particle motions on this Fock-space graph.  To that end let us compute the equation of motion for the position operator. It is useful to first list the algebraic relations among $D_\mu$, $V_\mu$, and $X_\mu$ operators,
 \begin{equation} \label{eqn:VDAlgebra}
      \begin{split}
 &    D_\mu   D_\nu = (\cos F_{\mu\nu})  D_\nu  D_\mu -(i \sin F_{\mu\nu})  V_\nu V_\mu,\\
& V_\mu V_\nu = (\cos F_{\mu\nu} )V_\nu V_\mu-(i \sin F_{\mu\nu} )  D_\nu  D_\mu,\\
  &  D_\mu V_\nu = (\cos F_{\mu\nu} )V_\nu  D_\mu+(i \sin F_{\mu\nu} )  D_\nu V_\mu, \\
 & V_\mu  D_\nu = (\cos F_{\mu\nu} ) D_\nu V_\mu+(i \sin F_{\mu\nu} ) V_\nu  D_\mu, \\
  &[ D_\mu, X_\nu] = [ V_\mu, X_\nu] = [ X_\mu, X_\nu]=0,
 \end{split}
 \end{equation}
for $\mu\neq \nu$ and that
 \begin{equation}\label{eqn:VDAlgebra2}
      \begin{split}
& V_\mu  D_\mu =-D_\mu V_\mu= i \sigma_\mu^3 =2i X_\mu,\\
& V_\mu X_\mu= - X_\mu V_\mu=-i D_\mu/2,\\ 
& D_\mu X_\mu= - X_\mu D_\mu=i V_\mu/2, \\
&    D_\mu^2= V_\mu^2 = (2X_\mu)^2= \mathbb{1}.
\end{split}
 \end{equation}
Among other things this shows $\{D_\mu\}$, $\{V_\mu\}$ and $\{X_\mu\}$ form an algebra under multiplication and addition.  We can also augment this alegbra by including  $\tilde D_\mu$ and $\tilde V_\mu$, and use 
\begin{align} \label{eq: T_commutation}
     T_\mu^\pm \tilde  T_\nu^\pm = \tilde  T_\nu^\pm  T_\mu^\pm e^{i\frac{F_{\mu\nu}+\tilde{F}_{\mu\nu}}{2}},\   T_\mu^\pm \tilde  T_\nu^\mp = \tilde  T_\nu^\mp  T_\mu^\pm e^{-i\frac{F_{\mu\nu}+\tilde{F}_{\mu\nu}}{2}}.
\end{align}
The end result is essentially to replace all the $F_{\mu\nu}$ in equation \eqref{eqn:VDAlgebra} by an average $(F_{\mu\nu}+\tilde F_{\mu\nu})/2$, for example
 \begin{equation} \label{eqn:VDAlgebraMixed}
      \begin{split}
 &    D_\mu   \tilde D_\nu = \left(\cos \frac{F_{\mu\nu}+\tilde F_{\mu\nu}}{2}\right)  \tilde D_\nu  D_\mu - \left(i \sin \frac{F_{\mu\nu}+\tilde F_{\mu\nu}}{2}\right)   \tilde V_\nu V_\mu,\\
  &  D_\mu \tilde V_\nu =\left(\cos \frac{F_{\mu\nu}+\tilde F_{\mu\nu}}{2}\right)  \tilde V_\nu  D_\mu + \left(i \sin \frac{F_{\mu\nu}+\tilde F_{\mu\nu}}{2}\right)   \tilde D_\nu V_\mu, \\
 \end{split}
 \end{equation}
As we have seen in equation
\eqref{eqn:DasVelocity} that $\dot X_\mu = V_\mu$, now let us derive a ``Newtonian'' (second derivative in time) equation in the following way:
\begin{equation}\label{eqn:EOM}
    \begin{split}
         &\dot X_\mu  D_\mu =  V_\mu  D_\mu =  2i X_\mu,\\
         \implies & \ddot X_\mu  D_\mu + \dot X_\mu \dot  D_\mu =2 i \dot X_\mu =2i V_\mu,\\ \implies & \ddot X_\mu  D_\mu^2 + \dot X_\mu \dot  D_\mu  D_\mu =2i V_\mu  D_\mu =-4X_\mu, \\
         \implies &\ddot X_\mu + \dot X_\mu \dot  D_\mu  D_\mu +4X_\mu = 0,
    \end{split}
\end{equation}
 where we have used  identities presented in equations \eqref{eqn:VDAlgebra} and \eqref{eqn:VDAlgebra2}. The harmonic term $4 X_\mu$ is solely due to the finiteness of the hypercube and is present even if the magnetic field is turned off. The second term in the last line is nonzero if and only if the magnetic field is nonzero,  so it must play the role of Lorentz force. To see this more clearly, we write the second term as 
 \begin{equation}
 \begin{split}
        \dot X_\mu \dot  D_\mu  D_\mu &=i V_\mu [H, D_\mu ]D_\mu = i \sum_{\nu \neq \mu}  V_\mu [D_\nu, D_\mu ]D_\mu\\
        &= i\sum_{\nu \neq \mu}  V_\mu D_\nu -i\sum_{\nu \neq \mu}  V_\mu D_\mu D_\nu D_\mu \\
        & = i\sum_{\nu \neq \mu}  (1-\cos F_{\mu\nu})V_\mu D_\nu -\sum_{\nu \neq \mu}  \sin F_{\mu\nu} D_\mu V_\mu \\
         & = i\sum_{\nu \neq \mu}  (1-\cos F_{\mu\nu})V_\mu D_\nu -D_\mu \sum_{\nu \neq \mu}  \sin F_{\mu\nu} \dot X_\nu,
 \end{split}
 \end{equation}
where we have used multiple identities in equations \eqref{eqn:VDAlgebra} and \eqref{eqn:VDAlgebra2}. So far all the equations above are exact. To see the ``continuum limit'' of the motion,\footnote{See the discussion below equation \eqref{eqn:schwarzianOpDim}.} we take a small $F_{\mu\nu}$ and Taylor expand equation \eqref{eqn:EOM} to the first power of $F_{\mu\nu}$ with fixed $d$, and this gives
\begin{equation}\label{eqn:truncateLorentz}
        \dot X_\mu \dot  D_\mu  D_\mu = -\sigma^1_\mu \sum_{\nu} F_{\mu\nu}\dot X_\nu + O(F_{\mu\nu}^2),
 \end{equation}
where we used $D_\mu (t) = \sigma^1_\mu +O(F_{\mu\nu})$. Hence in the continuum limit the Newtonian equation becomes 
\begin{equation}\label{eqn:lorentzForce}
\ddot X_\mu - \sigma^1_\mu \sum_{\nu} F_{\mu\nu}\dot X_\nu +4X_\mu = 0.
\end{equation}
    The second term is almost what one expects from a classical Lorentz force except here we have an extra $\sigma_\mu^1$ coefficient. The $\sigma_\mu^1$ coefficient seems to suggest there is no truly classical Lorentz force on the hypercube.\footnote{The conventional continuum limit  would have a shrinking lattice spacing $a$ with some length scale $L = n a$ fixed, we can repeat the same derivation of the equation of motion and the upshot is the replacement of $\sigma^1$ by $ a\Sigma^1 + 1$ where $\Sigma^1$ is the hopping on the large lattice and $1$ comes from the fact that $D_\mu$ needs a constant counter-term to become a well-defined continuum derivative. As $a \to 0$, we get the conventional Lorentz force.} In fact, without the $\sigma_\mu^1$ coefficient we will be in trouble: the motion becomes exactly the same as a harmonic oscillator in a uniform magnetic field, which is integrable, contradicting the chaotic physics of the Parisi model we have known so far. A useful comparison is a $d=2$ lattice with a large linear size,  which is a Landau problem for Bloch electrons whose spectrum is known to be complicated but integrable \cite{WIEGMANN1994495}.
Note that the Lorentz force term is a sum over $d$ terms, so it could dominate over the harmonic potential even when each $F_{\mu\nu}$ is very small. This is consistent with our earlier result that the system is fast scrambling when $F_{\mu\nu}$ is sent to zero after the large $d$ limit is taken. Moreover, according to a numerical analysis of the Parisi model \cite{jia_unpublished}, it indeed looks like any nonzero small magnetic field would lead to random matrix statistics and hence quantum chaos.  It should be said that if we want to take the large $d$ limit,  the truncation to linear order in \eqref{eqn:truncateLorentz} is not valid unless $\vev {F^2}\ll 1/d^2$ because the $O(F^2)$ term involves summing $d^2$ terms.  Nevertheless equation \eqref{eqn:lorentzForce} could be useful; for example, it  gives us a hint on what kind of scaling  a localizing potential must have so that our model can have a localized phase. For example, we can add to the Parisi Hamiltonian a term 
\begin{equation}
 - \sum_\mu  \mathcal{E}_\mu \frac{\sigma^3_\mu}{2}= -\sum_\mu  \mathcal{E}_\mu X_\mu
\end{equation}
where $ \mathcal{E}_\mu$ can be interpreted as the components of an electric field acting on the charged particle, and hence has the effect of localizing the particle on hypercube vertices.  This electric term would change the equation of motion \eqref{eqn:lorentzForce} to 
\begin{equation}
\ddot X_\mu - \sigma^1_\mu \sum_{\nu} F_{\mu\nu}\dot X_\nu +4X_\mu - \sigma^1_\mu \mathcal{E}_\mu= 0.
\end{equation}
Since the Lorentz force term sums over $d$ random numbers, we estimate its magnitude to be roughly $\sqrt{d \vev{F_{\mu\nu}^2}}$. Therefore $\mathcal{E}_\mu$ must at least scale as $\sqrt{d \vev{F_{\mu\nu}^2}}$ for it to be a real competition with the Lorentz force term. This is naturally satisfied in a scaling regime that has an extensive free energy (see section \ref{sec:extensiveScalings}):
\begin{equation}
   \vev{F_{\mu\nu}^2}\sim \frac{1}{{d}},\quad \vev{\mathcal{E}_\mu^2}\sim 1.
\end{equation}
However, this scaling is much less understood.

\section{Subleading moments at $q=1$ and $q=0$}\label{app:subleadingq1and0}
The subleading moments involve two pieces of contributions:  One is from the $d(d-1)\cdots$ factor of the leading lattice paths,  which (for $M_{2p}$) is equal to 
\begin{equation}
-\frac{1}{d}\binom{p}{2} M^{\rm QH}_{2p},
\end{equation}
 where $M^{\rm QH}_{2p}$ denotes the $q$-Hermite moments. The other is from the subleading lattice paths (with repeatedly traversed lattice directions),  whose values can be computed by equation \eqref{eqn:trueSubleading} (the same computation was called the merge and delete procedure of intersection graphs  in \cite{Jia_2020}).  At $q=1$ and $q=0$ subleading lattice path contributions can be figured out for general $p$,
 \begin{align}
 \frac{1}{d}\frac{1}{3}\binom{p}{2}(2p-1)!!   \quad &\text{for $q=1$,} \\ 
 \frac{1}{d}\binom{2p}{p-2}\frac{p}{2}  \quad &\text{for $q=0$.} 
 \end{align}
Hence the total subleading corrections at $q=1$ and $q=0$ are\footnote{We use the fact that $ M^{\rm QH}_{2p}(q=1) =(2p-1)!!$ and $M^{\rm QH}_{2p}(q=0) =\frac{1}{p+1}\binom{2p}{p}$.} 
 \begin{align}
\delta M_{2p}=& -\frac{1}{d}\frac{2}{3}\binom{p}{2}(2p-1)!!   \quad \text{for $q=1$,} \\ 
\delta M_{2p}=& - \frac{1}{d}\binom{2p}{p-2}  \quad \text{for $q=0$,} 
 \end{align}
 from which we can work out the subleading corrections to spectral densities,
 \begin{align}
\delta\rho =& \frac{1}{d}   \frac{d^4}{dx^4}\left[ -\frac{1}{12}  \frac{e^{-\frac{x^2}{2}}}{\sqrt{2\pi}}\right] \quad \text{for $q=1$,} \\
\delta \rho=&\frac{1}{d} \frac{d^2}{dx^2}\left[ \frac{1}{120\pi} (4-x^2)^{\frac{3}{2}}(2-3x^2)\right] \quad \text{for $q=0$.} 
 \end{align}
For $q=0$ the subleading density diverges rapidly near the spectral edges $x=\pm 2$, this density should be understood in the distributional sense,  
 \begin{equation}
 \int f(x) \delta\rho(x) dx = \frac{1}{d} \int f''(x) \left[ \frac{1}{120\pi} (4-x^2)^{\frac{3}{2}}(2-3x^2)\right]dx.
 \end{equation}
 
\section{Degenerate perturbation for the first excited states}\label{app:degenPerturb}
The general form of equation \eqref{eqn:2mixing} is
\begin{equation}
\begin{split}
    \langle m,n| \sqrt{d} H |0\rangle =&\sum_{\mu\neq m,n} \left[\sin \left(\frac{F_{\mu m}}{4}\right)\sin \left(\frac{F_{\mu n}}{4}\right) \prod_{\nu \neq\mu,m,n} \cos \left(\frac{F_{\mu \nu}}{4}\right) \right] \\
   &\quad +i \sin \left(\frac{F_{m n}}{4}\right) \left[ \prod_{\nu \neq m,n} \cos \left(\frac{F_{m\nu}}{4}\right)-\prod_{\nu \neq m,n} \cos \left(\frac{F_{n\nu}}{4}\right)\right].
    \end{split}
\end{equation}
Matrix elements of the form $\langle m | \sqrt{d}H |n\rangle$ are also simple to compute. Using the notations of equation \eqref{eqn:completSetH0},  we get the matrix elements of $H$ using the wave functions of the first excited states ($d$ states) of $H_0$:
\begin{equation}
    \begin{split}
        \langle m |\sqrt{d}H| m\rangle = -(d-2)\cos^{d-1}\left(\frac \phi 4\right),
    \end{split}
\end{equation}
and the off-diagonal elements ($m \neq n$) are 
\begin{equation}
    \langle m |\sqrt{d} H | n \rangle = \cos^{d-3}\left(\frac \phi 4\right) \sin^2\left(\frac \phi 4\right) \sum_{\mu\neq m,n}S_{\mu m}S_{\mu n}-2 i \cos^{d-2}\left(\frac \phi 4\right) \sin\left(\frac \phi 4\right)S_{mn}
\end{equation}
for  Parisi's disorder distribution. In full generality, we have 
\begin{equation}
    \begin{split}
        \langle m |\sqrt{d}H| m\rangle &= -\sum_{\mu\neq m}\prod_{\nu\neq\mu}\cos\left(\frac {F_{\mu\nu}} 4\right)+ \prod_{\nu\neq m}\cos\left(\frac {F_{m\nu}} 4\right)\\
        \langle m |\sqrt{d} H | n \rangle &= \sum_{\mu\neq m,n}\left[\sin\left(\frac{F_{\mu m}}{4}\right)\sin\left(\frac{F_{\mu n}}{4}\right)\prod_{\nu \neq m,n,\mu}\cos\left(\frac{F_{\mu \nu}}{4}\right)\right] \\
        &\quad \quad -i\sin\left(\frac{F_{m n}}{4}\right) \left[\prod_{\nu \neq m,n}\cos\left(\frac{F_{m \nu}}{4}\right)+\prod_{\nu \neq m,n}\cos\left(\frac{F_{n \nu}}{4}\right)\right], \quad m\neq n,
    \end{split}
\end{equation}
We can then numerically diagonalize the $d\times d$ matrix above and its lowest energy will approximate the first excited state energy of $H$. However, we did not need this in our main text.
\section{State mixing in the Maldacena-Qi model}\label{app:MQmixing}
In this appendix we demonstrate that the unperturbed ground state only mixes with a small fraction of the excited states in the MQ model. The MQ Hamiltonian is of the form
\begin{equation}
    H_{\rm MQ} = H_{\rm SYK, L}+ H_{\rm SYK, R} + i\mu \sum_{k=1}^d \psi_L^k \psi_R^k.
\end{equation}
The $\psi_L^k$ and  $\psi_R^k$ operators are two sets of Majoranas defined on the left and right spatial sites with the anticommutation relation
\begin{equation}
   \{\psi_a^k, \psi_b^l\} = 2\delta_{ab}\delta^{kl},
\end{equation}
where $a, b =L,R$ and $k,l =1,\ldots, d$.  $H_{\rm SYK, L}$ is the standard $p$-body SYK Hamiltonian built from the  $\psi_L$, and $H_{\rm SYK, R}$ is built from $\psi_R$  but with a time reversal.  As stated in section \ref{sec:spectralGap}, the coupling term is identical to the zero-flux Parisi Hamiltonian, so to parallel the discussion in the main text, we treat $ H_{\rm SYK, L}+ H_{\rm SYK, R} $ as a perturbation and study the mixing between the ground state and the excited states of the coupling term under this perturbation. Following \cite{garcia2019b}, we define ladder operators to generate the eigenstates of the coupling term,
\begin{equation}
    c^{\pm}_k =\frac{1}{\sqrt{2}} \left(\psi_R^k \pm  i\psi_L^k\right),
\end{equation}
then 
\begin{equation}
    i\sum_{k=1}^d \psi_L^k \psi_R^k = \sum_{k=1}^d c^+_k c^-_k -d 
\end{equation}
and the commutator
\begin{equation}
   \left [i\sum_{k=1}^d \psi_L^k \psi_R^k, \ c_l^{\pm} \right]= \pm 2  c_l^{\pm}.
\end{equation}
Hence we can label the eigenstates in the identical way as we did for the zero-flux Parisi Hamiltonian in equation \eqref{eqn:completSetH0}, now with 
\begin{equation}
    |m_1, \ldots,m_k\rangle := c^+_{m_1} \ldots  c^+_{m_k}|0\rangle,
\end{equation}
where the ground state $|0\rangle$ satisfies 
\begin{equation}
    c_k^- |0\rangle = 0
\end{equation}
for all $k=1,\ldots, d$. Now consider the mixing 
\begin{equation}
\begin{split}
   & \langle m_1, \ldots,m_l| H_L|0\rangle \\
   =&\langle 0 | c^-_{m_1} \ldots  c^-_{m_l} H_L|0\rangle \\
    =& \sum_{i_1<\ldots<i_p} J_{i_1\ldots i_p} \langle 0| c^-_{m_1} \ldots  c^-_{m_l} \psi_L^{i_1} \ldots \psi_L^{i_p}  |0\rangle.
\end{split}
\end{equation}
If $l>p$, there is at least one annihilation operator that anticommutes with every fermion in any given  set of $\{\psi_L^{i_1}, \ldots ,\psi_L^{i_p} \}$. Then, we can just move this annihilation operator all the way to the right and annihilate $|0\rangle$.  This means the ground state only mixes with excited states up to $|m_1, \ldots, m_p\rangle$, which only includes a vanishingly small fraction of all states as $d\to \infty$.
\bibliographystyle{JHEP}
\bibliography{library}

\end{document}